\documentclass[11pt,a4paper]{article}

\usepackage[margin=1in]{geometry}
\usepackage[authoryear,longnamesfirst]{natbib}

\usepackage{amsmath,amssymb,amsthm}
\usepackage{algpseudocode}
\usepackage{algorithm}
\usepackage{graphicx}
\usepackage{subcaption}
\usepackage{booktabs,multirow}
\usepackage[shortlabels]{enumitem}
\usepackage[english]{babel}
\usepackage{xcolor}
\usepackage{url}
\usepackage[colorlinks=true,citecolor=blue,linkcolor=blue,urlcolor=blue]{hyperref}
\newcommand{\tblwidth}{\textwidth}

\newcommand{\jia}[1]{}
\newcommand{\jial}[1]{}
\newcommand{\rich}[1]{}
\newcommand{\richl}[1]{}
\newcommand{\nicola}[1]{}
\newcommand{\nicolal}[1]{}

\title{Exploring Pareto smoothing in sequential Monte Carlo}

\author{
Jia Le Tan\thanks{Corresponding author. Email: \texttt{jia-le.tan@warwick.ac.uk}.}
\and
Nicola D. Walker
\and
Richard G. Everitt
}

\date{
\small
Department of Statistics, University of Warwick, Coventry, United Kingdom\\
Centre for Environment, Fisheries and Aquaculture Science, Lowestoft, United Kingdom
}

\begin{document}

\maketitle

\begin{abstract}
A popular technique for reducing the variance of importance sampling (IS) estimators is to modify the weights of some importance points. One approach is to truncate the largest weights, which reduces variance but can introduce substantial bias. Pareto smoothed importance sampling (PSIS), by contrast, reduces the variance of the weights by fitting a generalised Pareto distribution to the upper tail of the weight distribution and replacing the weights in this tail with the corresponding expected quantiles from the fitted distribution. PSIS can therefore also reduce variance, but typically with less bias, and has been used successfully in IS-based approximations for Bayesian cross-validation. This paper explores the use of PSIS steps within sequential Monte Carlo (SMC) samplers, with a particular focus on approximate Bayesian computation (ABC)-SMC algorithms, where we aim to use Pareto smoothing to reduce the use of Markov chain Monte Carlo (MCMC) moves, each of which requires simulation from a model that is often computationally expensive. Our empirical investigation suggests that there are only minimal benefits to using Pareto smoothing in SMC, since the variance reduction through using a sequence of targets dominates the impact of the weight adjustment.
\end{abstract}

\noindent\textbf{Keywords:}
Pareto smoothed importance sampling;
sequential Monte Carlo;
approximate Bayesian computation;
importance sampling;
simulation-based inference.

\vspace{1em}

\section{Introduction}

\subsection{Pareto smoothed importance sampling}

\subsubsection{Importance sampling}

Monte Carlo methods provide a framework for approximating expectations \(\mathbb{E}_{\pi}[g(\theta)]\) with respect to a target distribution \(\pi\) based on random samples. In particular, if \(\theta^{1:N} \sim \pi\), then the expectation
\[
\mathbb{E}_{\pi}[g(\theta)] = \int g(\theta)\,\pi(\theta)\,d\theta
\]
can be approximated by the Monte Carlo estimator
\[
\hat{\mathbb{E}}_{\pi}^{\mathrm{MC}}[g(\theta)] = \frac{1}{N}\sum_{i=1}^{N} g(\theta^{i}).
\]
When direct sampling from \(\pi\) is not feasible, importance sampling (IS) provides a general framework for approximating such expectations by using samples \(\theta^{1:N}\sim q\) drawn from a proposal distribution \(q\). Throughout, we use
the notation \(\tilde{\cdot}\) to denote an unnormalised quantity:
\(\tilde{\pi}\) for an unnormalised version of the target and \(\tilde{w}\)
for an unnormalised weight. In IS, each point \(\theta^{i}\) is assigned
an unnormalised weight \(\tilde{w}^{i}=\tilde{\pi}(\theta^{i})/q(\theta^{i})\),
which is then normalised to give the weight \(w^{i}=\tilde{w}^{i}/\sum_{k=1}^{N}\tilde{w}^{k}\).
The population of weighted points provides an empirical distribution
approximating the target \(\pi\). Taking the expectation of \(g\) under
this empirical distribution yields the ``self-normalised'' IS estimator
\[
\hat{\mathbb{E}}_{\pi}[g(\theta)]=\sum_{i=1}^{N}w^{i}g(\theta^{i}).
\]
The performance of IS depends on the distance between the distributions
\(\pi\) and \(q\); \citet{agapiou2017intrinsic} make links to both the Kullback--Leibler divergence and the \(\chi^2\) divergence. When \(q\) is not close to \(\pi\), which is possible
for a poorly chosen \(q\) in low-dimensional \(\theta\) and is practically
guaranteed as the dimension of \(\theta\) increases, IS suffers from
weight degeneracy: most of the mass in the empirical approximation
is concentrated on a few points that have very large weights. This
results in high-variance estimators \citep{tokdar2010importance,elvira2021advances}.

\subsubsection{Weight modification}

\citet{ionides2008truncated}, \citet{vehtari2024psis}, and \citet{vehtari2017practical}
consider improving IS estimators by modifying the weights of IS points.
These modifications introduce bias, but can significantly reduce
variance. \citet{ionides2008truncated} propose using modified
(unnormalised) weights, where the \(i\)th weight is given by
\begin{equation} \label{eq:truncation}
\min\left(\tilde{w}^{i},\frac{1}{\sqrt{N}}\sum_{i=1}^{N}\tilde{w}^{i}\right),
\end{equation}
thereby truncating weights that are large. This scheme ensures finite-variance
estimators when \(\mathbb{E}\!\left[\left|\tilde{w}^{i}\right|\right] < \infty\),
compared to the condition \(\mathbb{E}\!\left[\left|\tilde{w}^{i}\right|^{2}\right] < \infty\)
required for standard IS. \citet{vehtari2024psis} propose
an alternative scheme that also guarantees finite variance under weaker
conditions than standard IS, but which has lower bias in situations
of practical interest. This approach, Pareto smoothed importance sampling (PSIS), fits a generalised Pareto distribution (GPD) to the upper tail of the observed weights and replaces those extreme values with their ``smoothed'' counterparts.

\subsubsection{PSIS} \label{sec:psis}

To apply Pareto smoothing (PS) to the unnormalised IS weights, we first
identify a suitable threshold \(a\) (such as a high quantile of the
weight distribution) and focus on the extreme weights \(\tilde{w}^{i}-a\),
for those \(\tilde{w}^{i}>a\). The GPD provides a parametric form for
modelling the distribution of extreme weights. Its density is given by
\[
p(y\mid a,\sigma,k)=\begin{cases}
\frac{1}{\sigma}\left(1+k\left(\frac{y-a}{\sigma}\right)\right)^{-\frac{1}{k}-1}, & k\neq0,\\[6pt]
\frac{1}{\sigma}\exp\left(-\frac{y-a}{\sigma}\right), & k=0,
\end{cases}
\]
with parameters \(\sigma>0\) and \(k\in\mathbb{R}\), defined for \(y\geq a\)
when \(k\geq0\), and for \(a\leq y\leq a-\sigma/k\) when \(k<0\). Once
we fit \((\hat{k},\hat{\sigma})\) to the extreme weights above \(a\), we
use the GPD's inverse CDF,
\[
F^{-1}(z)=a+\frac{\sigma}{k}\left((1-z)^{-k}-1\right),
\]
to produce smoothed values for the extreme weights.

After computing the unnormalised weights, we sort them and determine
the threshold \(a\). Let \(M\) be the number of extreme weights above \(a\).
\citet{vehtari2024psis} recommend setting \(a\) such
that \(M=\left\lfloor \min\left(0.2N,3\sqrt{N}\right)\right\rfloor\).
We can represent the sorted weights as \(\tilde{w}^{N-M+z}\), where \(z=1,\ldots,M\),
which are precisely the extreme weights. To smooth the tail, we replace each extreme
weight \(\tilde{w}^{N-M+z}\) by a value obtained from equally spaced
quantiles of the fitted GPD's inverse CDF, i.e.
\[
\tilde{w}_{N-M+z}=F^{-1}\left(\frac{z-1/2}{M}\right)=a+\frac{\hat{\sigma}}{\hat{k}}\left(\left(1-\frac{z-1/2}{M}\right)^{-\hat{k}}-1\right).
\]
This transformation ensures that the largest weights are shrunk towards
values consistent with the fitted GPD, reducing their extremity and,
hence, the variance in the posterior estimates.

The \(m\)th moment of the GPD exists if \(k<1/m\); thus, the estimated value \(\hat{k}\) of \(k\) can be used as a simple diagnostic of the likely properties of IS estimates based on the simulated points. In particular, \(\hat{k}>0.5\) indicates infinite variance, leading to unstable estimates from standard IS. Further, under certain conditions, \citet{vehtari2024psis} link \(k\) to the KL divergence between \(\pi\) and \(q\), suggesting that \(\hat{k}\) may be used as an indicator of the quality of \(q\) as an IS proposal for estimating properties of \(\pi\). \citet{vehtari2024psis} show that, whilst standard IS performs poorly for \(k>0.5\), PSIS often performs well for \(k\) up to 0.7. Empirical results in that paper suggest that PSIS may outperform IS for any value of \(k\), but that the improvement is usually most pronounced for \(k>0.5\).

PSIS has been used extensively in leave-one-out cross-validation for Bayesian
models \citep{vehtari2017practical}. \citet{vehtari2024psis}
provide further empirical exploration of its properties, along with conditions
for its asymptotic consistency and finite variance. They also show empirically
that PSIS estimators are likely to have low bias and low variance when the
estimated \(\hat{k}<0.7\).

\subsection{Sequential Monte Carlo} \label{sec:smc}

This paper performs an empirical investigation of the use of Pareto smoothing in sequential Monte Carlo (SMC) samplers \citep{delmoral2006smc}. One way of viewing SMC is as an alternative approach to improving IS, through the introduction of a sequence of intermediate distributions between \(q\) and \(\pi\). Here we focus on the variant that uses Markov chain Monte Carlo (MCMC) moves on the points at each iteration.

The algorithm begins by simulating points \(\theta_{0}^{1:N}\) from
\(\pi_{0}:=q\) and assigning them all the same weight \(w^{i}=1/N\)
for \(i=1:N\). Then, at the \(t\)th iteration, the points are reweighted
using an IS step:
\begin{equation} \label{eq:reweight}
\tilde{w}_{t}^{i}=w_{t-1}^{i}\frac{\tilde{\pi}_{t}\left(\theta_{t-1}^{i}\right)}{\tilde{\pi}_{t-1}\left(\theta_{t-1}^{i}\right)},
\end{equation}
the weights are normalised as \(w_{t}^{i}=\tilde{w}_{t}^{i}/\sum_{k=1}^{N}\tilde{w}_{t}^{k}\),
and the particles are then resampled and moved, which together can be summarised
by
\begin{equation}
\theta_{t}^{i}\sim\sum_{j=1}^{N} w_{t}^{j}K_{t}\left(\cdot\mid\theta_{t-1}^{j}\right),\qquad w_{t}^{i}=1/N,
\label{eq:resample-move}
\end{equation}
where \(K_{t}\) is an MCMC kernel with invariant distribution \(\pi_{t}\).

The variance of estimators based on SMC output can be controlled by
ensuring that the distance between successive targets is small, as
long as the MCMC kernels \(K_{t}\) mix well; see, for example, \citet{chopin2020smcbook} for a detailed treatment.

One of the advantages of using SMC in practice is that it can be tuned automatically as the algorithm is running. A key quantity in several of these adaptive schemes is the effective sample size (ESS) \citep{kong1994sequential} that quantifies weight degeneracy:
\begin{equation}
\mathrm{ESS}_t
=
\frac{\left(\sum_{i=1}^N \tilde w_t^i\right)^2}
{\sum_{i=1}^N (\tilde w_t^i)^2}.
\end{equation}
It can be seen \citep{agapiou2017intrinsic} to be an estimator of \(N/(D(\pi_t\mid \pi_r)+1)\), where \(D(\pi_t\mid \pi_r)\) is the \(\chi^2\)-divergence between the current target \(\pi_t\) and \(\pi_r\), the target at which resampling last took place. To control the variance of estimates from SMC, a commonly used idea is to choose the sequence of targets adaptively to achieve a reasonable trade-off between avoiding degeneracy at each iteration and the additional computational complexity of adding further targets. The most popular scheme is to choose the target at iteration \(t\) such that the ESS is some pre-determined proportion \(\beta\) of \(N\) \citep{delmoral2012adaptiveabc}. \citet{delmoral2006smc} show that, when using MCMC kernels, the weight update in equation (\ref{eq:reweight}) provides low-variance estimators as long as the sequence of targets does not change too quickly, so \(\beta\) is usually chosen to be close to 1 (0.9, for example) to ensure this.

The kernel \(K_{t}\) is also often chosen adaptively,
for example by using a Gaussian random-walk kernel with mean \(\theta_{t-1}^{i}\)
and variance taken to be a scaled version of the sample variance of
the current particle population. In many implementations, the resampling
step is not used at every iteration, and is instead triggered adaptively
when the ESS falls below some threshold (often \(0.5N\)). When the MCMC move is particularly expensive, as in the SMC\(^2\)
algorithm \citep{chopin2013smc2}, an effective trade-off between statistical
and computational efficiency may be achieved by implementing
the move step only when resampling is triggered, i.e. by using the step
in equation (\ref{eq:resample-move}) only when the ESS falls below some threshold.

\subsection{Outline of the paper and previous work}

In section \ref{sec:algorithms} we introduce the algorithms studied in the paper, followed by a study of the following different aspects of using Pareto smoothing in SMC:

\begin{itemize}

\item The relationship of the \(\hat{k}\) diagnostic to the KL divergence between the IS proposal and target suggests that it may be of use in automatically tuning the sequence of targets used in SMC. To investigate its use in this situation, and to understand when PS may be of use in SMC, we study the link between \(\hat{k}\) and the ESS (section \ref{sec:ESSkhat}).

\item We examine the effect of using PS in SMC in section \ref{sec:ps-smc}.

\item One way of thinking about Pareto smoothing is as a method for mitigating degeneracy. We assess its effectiveness in mitigating degeneracy in SMC as an alternative to other approaches, also in section \ref{sec:ps-smc}. In particular, in section \ref{sec:abc} we assess its effectiveness in reducing the use of computationally expensive MCMC moves in approximate Bayesian computation (ABC).

\end{itemize}

Weight modification schemes have been used previously in SMC-style algorithms. \citet{koblents2015pmc} and \citet{vazquez2017transformed} investigate the use of annealed or truncated importance weights in the context of population Monte Carlo (PMC). Like SMC, PMC is a sequential IS method. The target distribution in PMC is the same at each IS step, with the collection of importance points at iteration \(t-1\) being used solely to construct the proposal at iteration \(t\). In contrast, the SMC samplers in this paper use a sequence of targets in which the target from iteration \(t-1\) is effectively used as the proposal for iteration \(t\). In this case, there is the potential for bias from weight modification in the weight update at each step to accumulate across SMC iterations. \citet{everitt2017bayesian} show that, under strong mixing conditions on the kernel \(K_t\), a bias at each weight update does not accumulate unboundedly, but advise caution in the practical use of such a scheme.

The idea of using Pareto smoothing as an alternative to MCMC was introduced in \citet{burkner2020approximate}, and used as a mitigation for degeneracy in SMC in \citet{han2025adaptive}, where the application is cross-validation for Bayesian models. We study the adaptive approach used in this paper in section \ref{sec:smcvariants}.

\section{Pareto smoothing in sequential Monte Carlo}

\subsection{Algorithms} \label{sec:algorithms}

\subsubsection{General framework} \label{sec:ps-smc-algs}

SMC produces high-variance estimators when the particle population becomes degenerate. A natural idea is to use PS to mitigate the effects of degeneracy. For SMC with a fixed sequence of targets, we might hope that this reduces degeneracy at each iteration. When the sequence is chosen adaptively, we might hope that fewer intermediate targets are chosen without significantly increasing the error in the estimates.

Algorithm~\ref{alg:generic_pssmc} describes the SMC sampler used in this paper. Here, the PS step uses the method described in Section~\ref{sec:psis} on the unnormalised weights \(\tilde{w}^{1:N}_t\) at the \(t\)th iteration. We denote by \(\hat{a}_t\), \(\hat{\sigma}_t\), and \(\hat{k}_t\) the estimates at iteration \(t\) of the GPD parameters.

We present a single algorithmic template (Algorithm~\ref{alg:generic_pssmc}) that unifies the methods discussed above. Boolean parameters \texttt{PARETO}, \texttt{RESAMPLE}, and \texttt{MCMC} control whether Pareto smoothing, resampling, and/or MCMC moves are used at each stage.

\begin{algorithm}
\caption{Pareto smoothing in SMC}
\label{alg:generic_pssmc}
\begin{algorithmic}[1]
\Require Distributions \(\{\pi_t(\theta)\}_{t=0}^T\), booleans \texttt{PARETO}, \texttt{RESAMPLE}, \texttt{MCMC}, transition kernels \(\{K_t\}\), number of particles \(N\).
\State \textbf{Initialise:}
\State Sample \(\theta_0^i \sim \pi_0\) for \(i=1,\ldots,N\), and set \(w_0^i = 1/N\).
\For{\(t=1 \to T\)}
    \State \textbf{Weight update:} For each \(i\),
    \[
    \tilde{w}_t^i \;=\; \tilde{w}_{t-1}^i \,\frac{\pi_t\bigl(\theta_{t-1}^i\bigr)}{\pi_{t-1}\bigl(\theta_{t-1}^i\bigr)}.
    \]
    \If{\texttt{PARETO}}
        \State \textbf{Pareto smoothing:} Fit a GPD to the largest weights in \(\{\tilde{w}_t^i\}_{i=1}^N\) and replace them with GPD-smoothed values (while also obtaining \(\hat{k}_t\)):
        \[
        \hat{k}_t, \tilde{w}_t^{1:N} \;=\; \mathrm{PSIS}(\tilde{w}_t^{1:N}).
        \]
    \EndIf
    \If{\(t=T\)}
        \State Normalise the weights to obtain \(w_T^{1:N}\) and terminate.
    \EndIf
    \If{\texttt{RESAMPLE}}
        \State \textbf{Normalise:}
        \[
        w_t^i \;\leftarrow\; \frac{\tilde{w}_t^i}{\sum_{j=1}^N \tilde{w}_t^j}.
        \]
        \State \textbf{Resample:} Draw \(N\) particles \(\{\theta_t^{i*}\}\) from \(\{\theta_{t-1}^i\}\) using probabilities \(\{w_t^i\}\). Set \(\tilde{w}_t^i = 1/N\) for all \(i\).
    \Else
        \State \(\theta_t^{i*} \leftarrow \theta_{t-1}^i\).
    \EndIf
    \If{\texttt{MCMC}}
        \State \textbf{MCMC move:} For each \(i\),
        \[
        \theta_t^i \;\sim\; K_t\bigl(\,\cdot\,\mid\theta_t^{i*}\bigr).
        \]
    \Else
        \State \(\theta_t^i \leftarrow \theta_t^{i*}\).
    \EndIf
\EndFor
\end{algorithmic}
\end{algorithm}

The empirical investigations in this paper make different choices as to whether the different steps of the algorithm are executed at each iteration of the SMC.


When PS is never triggered, the algorithm corresponds to standard forms of SMC:

\begin{itemize}
    \item with no resampling or MCMC, the algorithm reduces to importance sampling with proposal \(q\) and target \(\pi\);
    \item with MCMC but no resampling, the algorithm is annealed IS \citep{neal2001ais};
    \item with resampling but no MCMC, all particles will end up being duplicates of the same point;
    \item using resampling and MCMC gives a standard SMC sampler with MCMC moves.
\end{itemize}


Situations in which we might expect PS to be beneficial in SMC are those in which the additional bias is not significant compared to the variance reduction due to PS. The usual means of reducing the variance in SMC is by increasing the computational effort employed, usually through increasing the number of particles \(N\) or the number of targets \(T\).




\subsubsection{Algorithms studied} \label{sec:smcvariants}
We use the algorithmic template from Algorithm~\ref{alg:generic_pssmc} to construct six variants that incorporate different combinations of Pareto smoothing, resampling, and MCMC moves into SMC samplers:
\begin{itemize}
    \item \emph{SMC sampler}: 
    \texttt{PARETO=False} and \texttt{RESAMPLE=MCMC=True}. This uses the standard approach of resampling and MCMC to combat degeneracy.
    \item \emph{Pareto smoothed sequential importance sampling (PS-SIS)}:
    \texttt{PARETO=True} and \texttt{RESAMPLE=MCMC=False}. Here we replace resampling and MCMC with Pareto smoothing of the weights at each iteration: we do not expect this to be an effective approach, but we use it to study the effect of using Pareto smoothing iteratively.
    \item \emph{Pareto smoothed SMC sampler (PS-SMC)}: 
    \texttt{PARETO=True} and \texttt{RESAMPLE=MCMC=True}. This is the most straightforward application of PS to SMC: simply using PS on the weights at each iteration. We use PS at every SMC step, regardless of whether there is an indication that the variance-reduction approach might be needed.
    \item \emph{Adaptive SMC sampler (a-SMC)}:
    \begin{itemize}
        \item If \(\hat{k} \le 0.7\), then \texttt{PARETO=False} and \texttt{RESAMPLE=MCMC=False}.
        \item If \(\hat{k} > 0.7\), then \texttt{PARETO=False} and \texttt{RESAMPLE=MCMC=True}.
    \end{itemize}
    This algorithm does not do anything as it moves through the targets until it
    detects that a weighting step will be close to degenerate, via finding that
    \(\hat{k}\) reaches a value above 0.7. At that point, it uses a standard
    SMC resample--move step. The end result is (approximately) that the sequence of targets is chosen adaptively using the criterion $\hat{k}\geq 0.7$ to choose each SMC target. The appeal of this approach is pointed out in \citet{vehtari2024psis}, referencing the recommendation of \citet{chatterjee2018sample} to use an IS diagnostic ``that is not itself an importance sampling estimate of any quantity'', and pointing out that \(\hat{k}\) is such a diagnostic. We have included a-SMC in the paper to give an insight as to whether this is a useful approach.
    \item \emph{Adaptive Pareto smoothed SMC sampler (aPS-SMC)}:
    \begin{itemize}
        \item If \(\hat{k} \le 0.7\), then \texttt{PARETO=False} and \texttt{RESAMPLE=MCMC=False}.
        \item If \(\hat{k} > 0.7\), then \texttt{PARETO=True} and \texttt{RESAMPLE=MCMC=True}.
    \end{itemize}
    This algorithm is the same as a-SMC, but with the addition of PS when doing the weight update, aiming to use PS to mitigate large gaps between targets.
    
    \item \emph{Adaptive Pareto smoothed SIS/SMC (aPS-SIS/SMC)}, introduced in \citet{han2025adaptive}:
    \begin{itemize}
        \item If \(\hat{k} \le 0.7\), then PS-SIS, i.e.\ \texttt{PARETO=True} and \texttt{RESAMPLE=MCMC=False}.
        \item If \(\hat{k} > 0.7\), then SMC, i.e.\ \texttt{PARETO=False} and \texttt{RESAMPLE=MCMC=True}.
    \end{itemize}
    In this method, iterative PS is used to mitigate degeneracy at each step (as in PS-SIS) until the weighting step is close to degenerate (via finding that \(\hat{k}\) reaches a value above 0.7), at which point it uses a resampling step and an MCMC move as in standard SMC. Again, the aim is to reduce the number of MCMC moves, using PS as a substitute when the degeneracy is not too extreme.
\end{itemize}

Several of the algorithms in this section are related to the approach in \cite{han2025adaptive}. The algorithm in that paper uses the ESS to tune the sequence of distributions, then uses the criterion \(\hat{k} = 0.7\) to decide whether to perform PS (\(\hat{k} < 0.7\)) or resampling and MCMC (\(\hat{k} \geq 0.7\)). In our paper we consider a sequence of distributions governed by a continuous tempering parameter, whereas in the cross-validation setup in \cite{han2025adaptive} the sequence is determined by the discrete number of data points used in the cross-validation. We have complete freedom to determine the sequence of distributions in our examples due to this continuous parameter, but this is not the case in the data-point tempering in \cite{han2025adaptive}. We believe this motivates the idea in that paper of using PS as mitigation. Our paper is not a critique of methods used in \cite{han2025adaptive}, rather it is an exploration of whether PS might be of use in SMC more generally.

\subsection{Relationship between \texorpdfstring{$\hat{k}$}{k-hat} and ESS}
\label{sec:ESSkhat}

The relationship between \(\hat{k}\) and the ESS is key to understanding how we may use PS in SMC. \citet{vehtari2024psis} established the benefits of using PS when $0.5<\hat{k}<0.7$, whereas the ESS is the most frequently used criterion for choosing sequences of targets in SMC. Recall from Section~\ref{sec:smc} that the ESS yields an estimator of the \(\chi^2\)-divergence between the current target \(\pi_t\) and \(\pi_r\), the target at which resampling last took place.


\citet{vehtari2024psis} establish links between the estimated Pareto shape
parameter \(\hat{k}\) and the ESS. Here, we empirically examine this
relationship in the same Gaussian settings used later in the paper, with the aims of: (a) assessing the relevance of \(\hat{k}\) in SMC; (b) gaining insight into whether PS might be used on sequences of targets typically employed in SMC.

We consider tempered paths
\begin{equation} \label{eq:tempered}
\pi_{\alpha}(x)\propto \pi_0(x)^{1-\alpha}\pi_1(x)^{\alpha},
\qquad \alpha\in[0,1],
\end{equation}
where \(\pi_0=N(0,1)\). We use three terminal distributions,
\begin{equation} \label{eq:gaussians}
\pi_1=N(10,1), \qquad
\pi_1=N(0,10^{-5}), \qquad
\pi_1=N(0,10),
\end{equation}
corresponding respectively to a mean shift, a variance contraction, and a
variance expansion.

For each setting, we simulate \(N=1000\) samples from \(\pi_0\), compute
importance weights with respect to the tempered target at a given value of
\(\alpha\), and repeat this over \(1000\) independent seeds. To trace the
empirical relationship between \(\hat{k}\) and ESS, we consider a fine grid of \(\hat{k}\) values in \([0,1]\). For each \(\hat{k}\) value in the grid, we use bisection
over \(\alpha\in[0,1]\) to find a tempered distribution for which the estimated
Pareto shape parameter is approximately equal to the desired value, and then
record the corresponding ESS fraction \(\mathrm{ESS}/N\), and for each seed, plot the resultant values of this fraction against \(\hat{k}\). 

\begin{figure}
    \centering

    \begin{subfigure}[t]{0.32\textwidth}
        \centering
        \includegraphics[width=\linewidth]{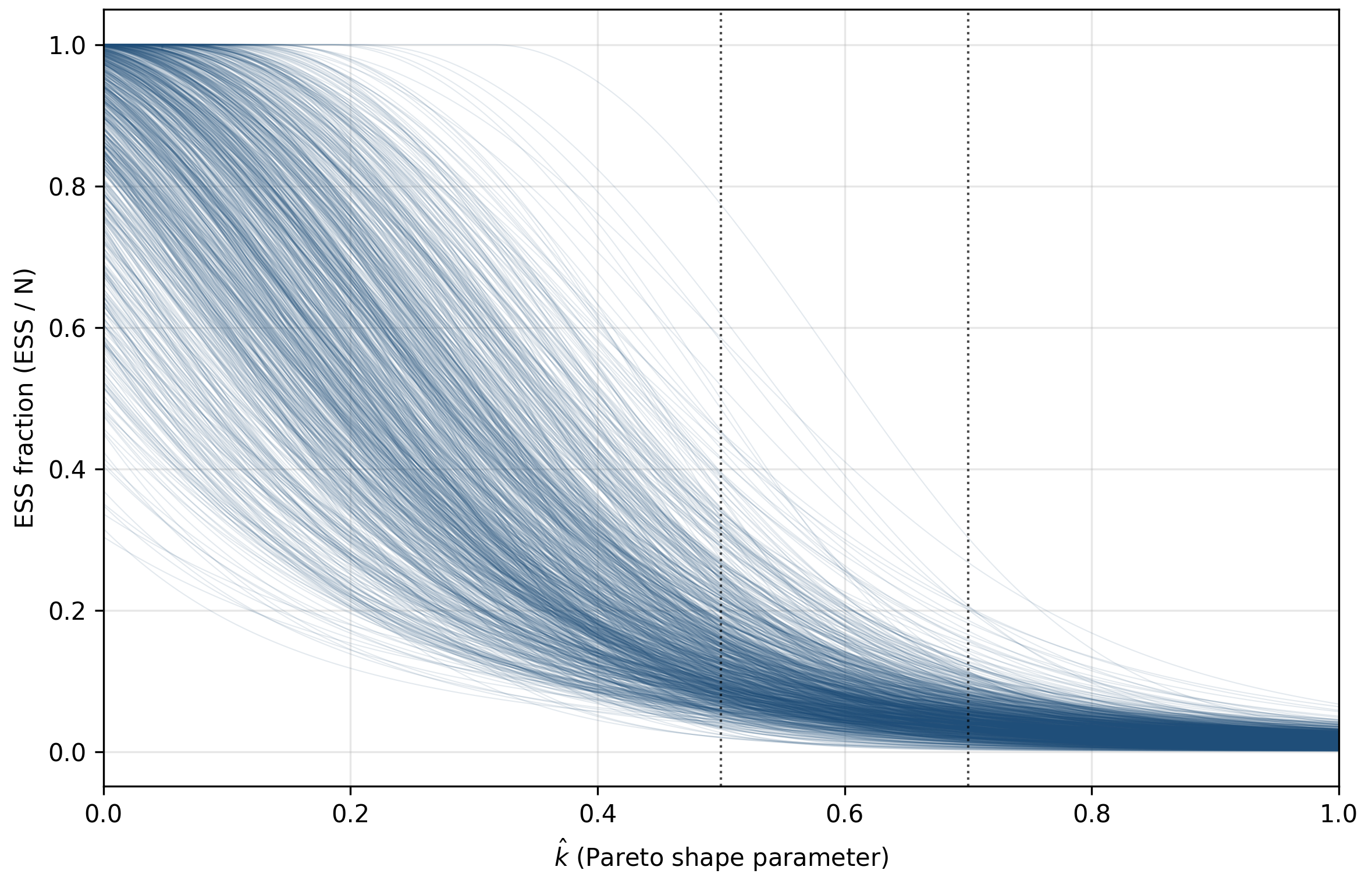}
        \caption{\(N(0,1)\rightarrow N(10,1)\).}
        \label{fig:ess-v-khat-mean-shift}
    \end{subfigure}
    \hfill
    \begin{subfigure}[t]{0.32\textwidth}
        \centering
        \includegraphics[width=\linewidth]{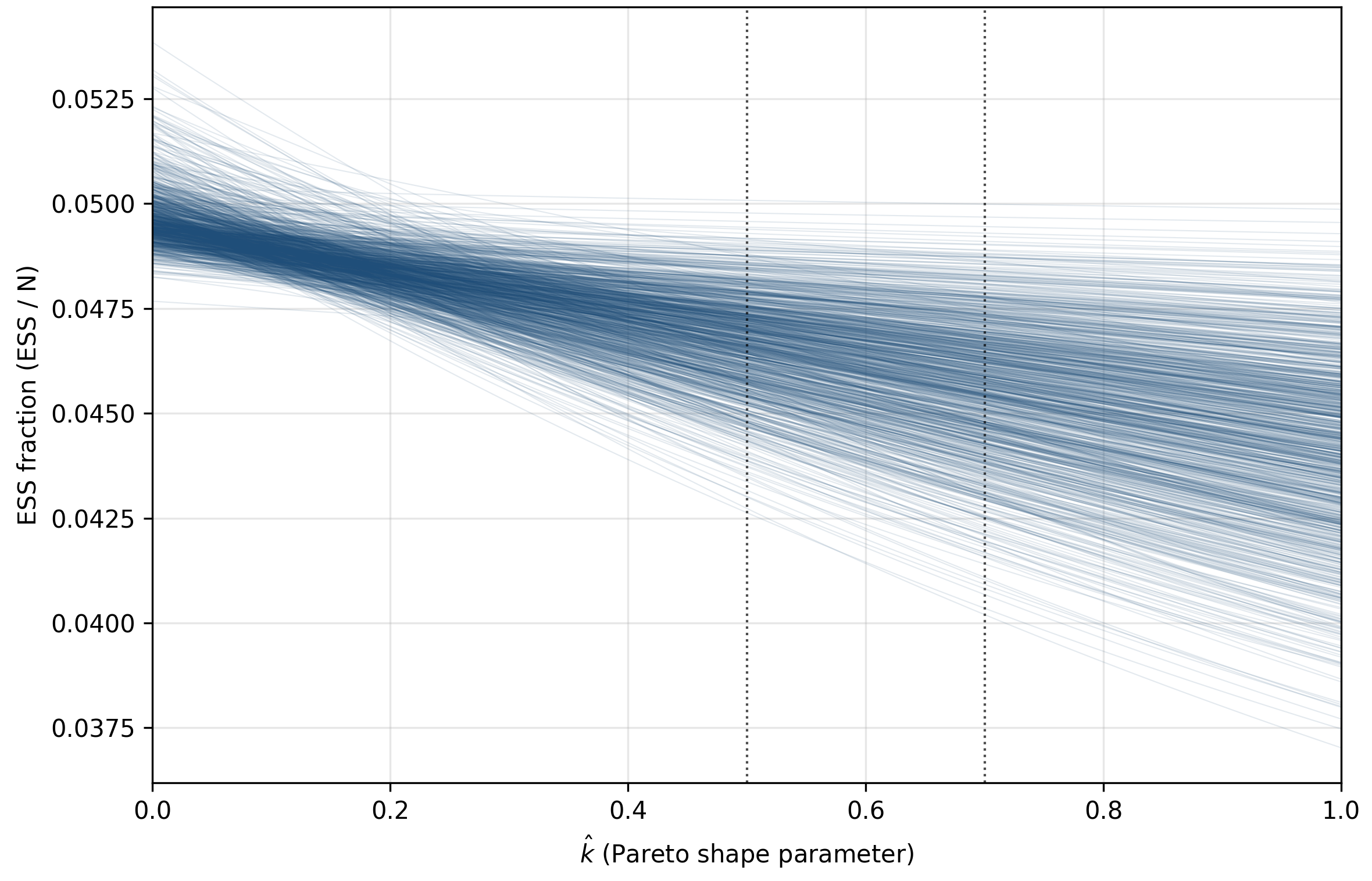}
        \caption{\(N(0,1)\rightarrow N(0,10^{-5})\).}
        \label{fig:ess-v-khat-thin}
    \end{subfigure}
    \hfill
    \begin{subfigure}[t]{0.32\textwidth}
        \centering
        \includegraphics[width=\linewidth]{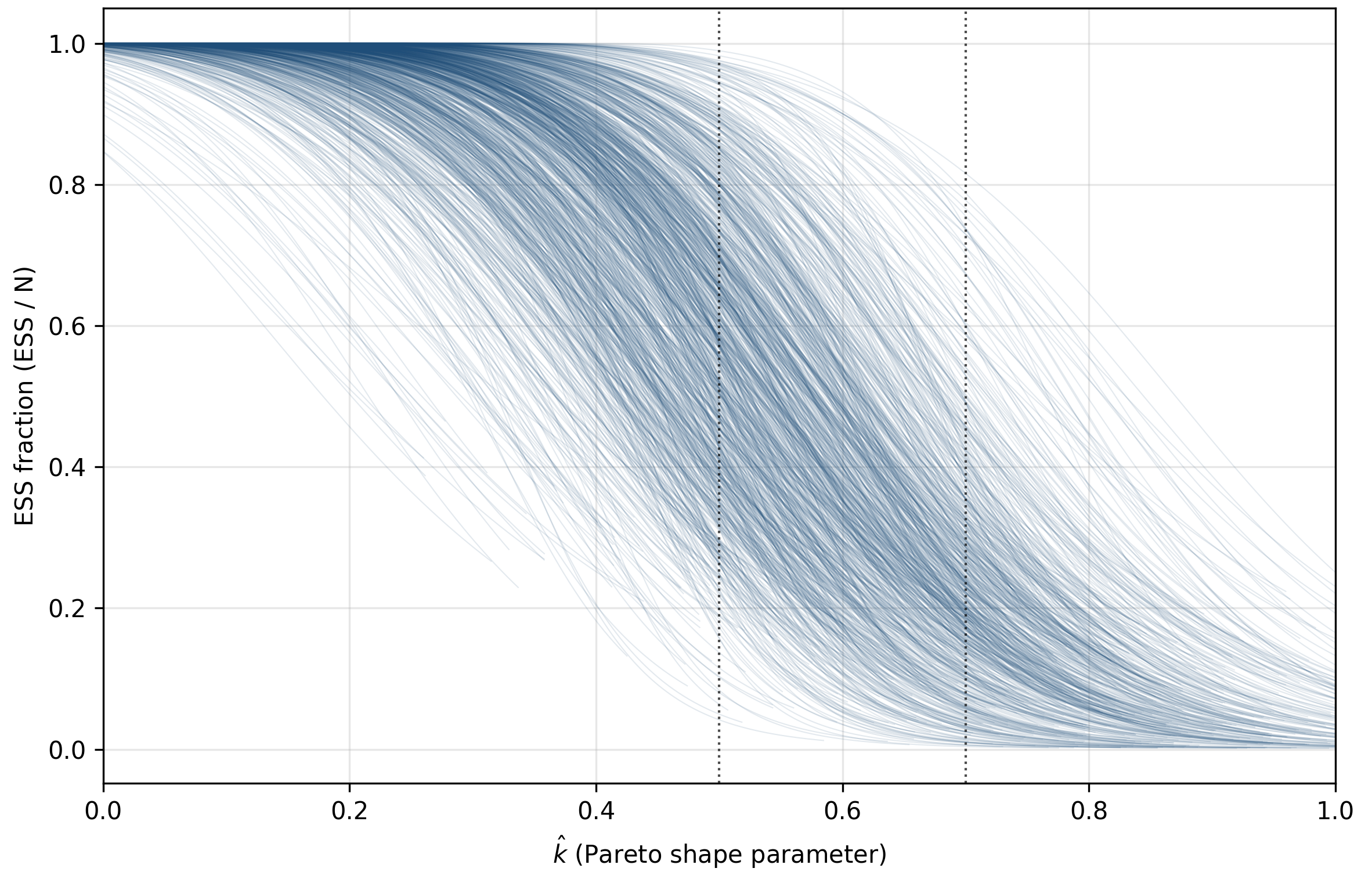}
        \caption{\(N(0,1)\rightarrow N(0,10)\).}
        \label{fig:ess-v-khat-fat}
    \end{subfigure}

    \caption{ESS fraction \(\mathrm{ESS}/N\) versus \(\hat{k}\) for three
    Gaussian examples. The proposal is \(N(0,1)\), and the targets are
    \(N(10,1)\), \(N(0,10^{-5})\), and \(N(0,10)\). Vertical dotted lines mark
    \(\hat{k}=0.5\) and \(\hat{k}=0.7\).}
    \label{fig:ess-v-khat}
\end{figure}

Recall that SMC targets are usually chosen so that the ESS does not drop too much between each target: choosing $\alpha$ such that the ESS is approximately $0.9N$ would be typical. We see that there is no guarantee that the region $0.5 < \hat{k} < 0.7$ in which PS was established to be beneficial might correspond to such a choice of targets. The situation where $\hat{k}$ is most likely to be in this region for a target sequence that might be used in SMC is the variance expansion setting in Figure~\ref{fig:ess-v-khat}(c). We might expect very little or no benefit in the variance contraction setting in figure \ref{fig:ess-v-khat}(b), where a sequence of targets chosen using the ESS would correspond to a negative $\hat{k}$ at each iteration.

For different proposals and targets, and different choices of $N$, we will observe a different relationship between the ESS and $\hat{k}$. Appendix \ref{app:ess_khat_extra} contains some additional examples of the relationship: using exponential targets and proposals as in \citet{vehtari2024psis}, for which the results are similar to those in figure \ref{fig:ess-v-khat}; and for a 100-dimensional version of the Gaussian example, where the graph in all cases (mean shift, variance contraction and expansion) is similar to figure \ref{fig:ess-v-khat}(a).

Since \(\hat{k}=0.7\) typically corresponds to small values of the ESS, we expect the a-SMC and aPS-SMC methods, in which targets are chosen so that \(\hat{k}\geq 0.7\), to usually result in large gaps between targets. Our experiments comparing a-SMC and aPS-SMC in the following section will assess whether this is mitigated by using PS.

\subsection{Gaussian example}
\label{sec:ps-smc}

\subsubsection{Experimental setup}

In this section we apply the algorithms from section \ref{sec:smcvariants} to the Gaussian examples from the previous section, i.e using the proposal $\pi_0=N(0,1)$, the three final targets in equation (\ref{eq:gaussians}) (mean shift, variance contraction and expansion), and the tempered sequence of targets in equation (\ref{eq:tempered}) for a sequence of powers $0 = \alpha_0 < ... < \alpha_T$.


The performance of an SMC sampler is highly dependent on the sequence $(\alpha_t)$. The different character of the three examples necessitates a different choice of sequence for each. For the three situations we consider, \citet{chopin2024tempering} uses ideas from mirror descent to analytically construct sequences of tempered distributions that are suitable for use in SMC (see Appendix~\ref{app:gaussian_tempering_schedule}). Using a sequence that is designed specifically for each situation allows some degree of comparison across the scenarios. This approach allows us to construct appropriate sequences with a pre-specified number of targets $T$.

For each value of \(T\), we run each method using \(N=1000\) particles and
compare the final weighted particle approximation with the true target
distribution using the one-dimensional Wasserstein distance, and by examining the mean square error of the first and second moment estimates. We vary \(T\) to
examine how performance changes as the path between \(\pi_0\) and \(\pi_T\)
is refined.

\subsubsection{Results}

\begin{figure}
    \centering

    \begin{subfigure}{\textwidth}
        \centering
        \includegraphics[width=\linewidth]{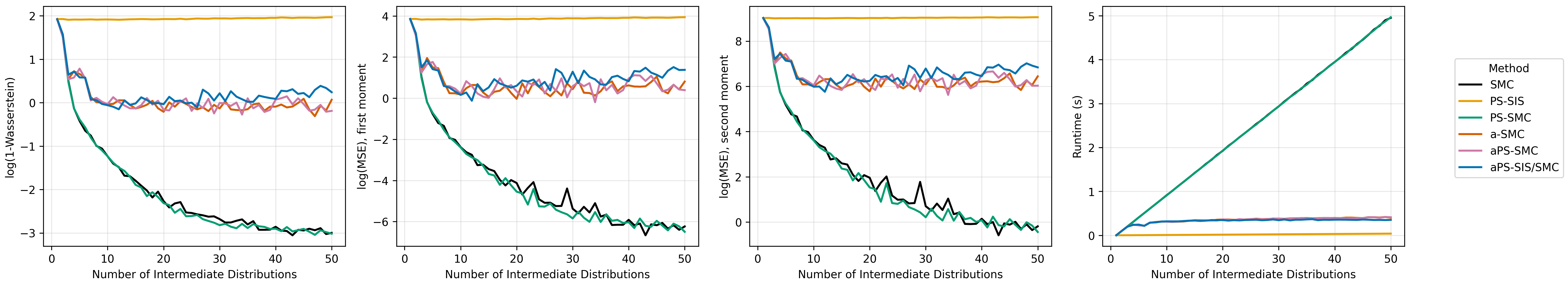}
        \caption{Mean-shift example:
        \(N(0,1)\rightarrow N(10,1)\).}
        \label{fig:gaussian_smc_results_mean_shift}
    \end{subfigure}

    \vspace{0.8em}

    \begin{subfigure}{\textwidth}
        \centering
        \includegraphics[width=\linewidth]{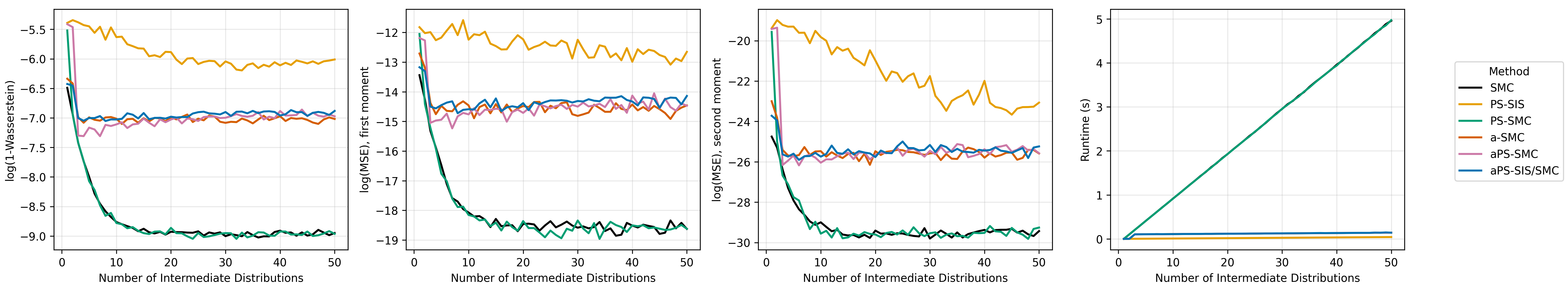}
        \caption{Variance-contraction example:
        \(N(0,1)\rightarrow N(0,10^{-5})\).}
        \label{fig:gaussian_smc_results_fat_to_thin}
    \end{subfigure}

    \vspace{0.8em}

    \begin{subfigure}{\textwidth}
        \centering
        \includegraphics[width=\linewidth]{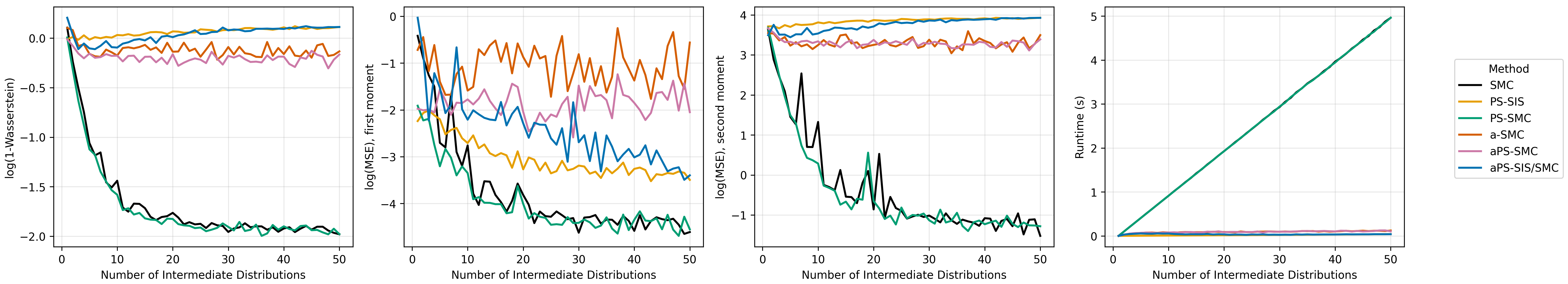}
        \caption{Variance-expansion example:
        \(N(0,1)\rightarrow N(0,10)\).}
        \label{fig:gaussian_smc_results_thin_to_fat}
    \end{subfigure}

    \caption{Performance of the SMC variants for the three one-dimensional
    Gaussian examples as the number of intermediate distributions \(T\) is
    varied. Panels correspond to the mean-shift example
    \(N(0,1)\rightarrow N(10,1)\), the variance-contraction example
    \(N(0,1)\rightarrow N(0,10^{-5})\), and the variance-expansion example
    \(N(0,1)\rightarrow N(0,10)\). Within each panel, the four subplots show
    the log Wasserstein distance, log mean-squared error for the first moment,
    log mean-squared error for the second moment, and average runtime. A more
    detailed decomposition into squared bias and variance is given in
    Appendix~\ref{app:gaussian_smc_extra}.}
    \label{fig:gaussian_smc_results}
\end{figure}

These experiments are designed to test three ideas: (a) using PS after a weight update in SMC; (b) using the criterion $\hat{k} = 0.7$ as an approach to adding additional targets in the SMC sequence; (c) using PS recursively as an approach to mitigate degeneracy.

The methods compared are standard SMC, PS-SIS, PS-SMC, a-SMC, aPS-SMC, and
aPS-SIS/SMC. Standard SMC performs resampling and MCMC moves at each
intermediate distribution. PS-SIS applies PS but does not
resample or move the particles. PS-SMC applies PS before the
usual resample--move step. The adaptive methods only perform resampling and MCMC moves when \(\hat{k}\) reaches 0.7, as in \citet{han2025adaptive} - recall from the previous section that this is likely to result in a relatively large gap between distributions in general. In particular, a-SMC is then simply standard SMC with a large gap between distributions. aPS-SMC uses a PS step each time resampling and MCMC are used, and aPS-SIS/SMC instead uses PS steps in place of resampling and MCMC moves when \(\hat{k}\leq 0.7\) as in \citet{han2025adaptive}.

Figure \ref{fig:gaussian_smc_results} shows the results. SMC and PS-SMC exhibit, as expected, improving performance as the number of target distributions increases, but also increasing computational cost. The performance of a-SMC and aPS-SMC changes little as the number of intermediate distributions increases after a certain point, since the criterion \(\hat{k}>0.7\) is not met at the additional intermediate targets.

We observe the following in relation to the ideas we are testing:

\begin{enumerate}[(a)]
    \item By comparing SMC to PS-SMC, and a-SMC to aPS-SMC, we see that PS only has a small impact compared to the improvement in performance we obtain by introducing more target distributions. The case where PS seems to have the clearest positive impact is in the variance expansion example, where the comparison of a-SMC to aPS-SMC highlights that there is a consistent small advantage in using PS when the gap between the distributions is relatively large.

    \item The error at which a-SMC stabilises gives us a guide as to what the error would be were the sequence of distributions chosen adaptively using $\hat{k}=0.7$. We see that in general the results are not positive: the error of the second moments in the mean-shift and variance-expansion examples is large. Whilst the idea of using $\hat{k}$ as a criterion by which to adapt the sequence of targets is attractive, it is not effective in practice since a path derived using this strategy has no guarantees of being close to the optimal schedule described in \citet{syed2024optimised}.

    \item The results of using PS recursively without resampling and SMC can be seen in the results for PS-SIS. The results for this approach are very poor: without MCMC moves, the bias introduced in the PS step accumulates across iterations. We also see this issue in some of the results for aPS-SIS/SMC, suggesting that if it is used at all, PS should only be used to adjust weights directly before resampling and MCMC.
    
\end{enumerate}

\subsection{Discussion}

The results above suggest that Pareto smoothing has limited value when the standard SMC resample--move mechanism is already computationally affordable. In such settings, the most reliable way to maintain accuracy is to move particles through the sequence of intermediate distributions. However, the trade-off might be different when move steps are expensive. In that case, even a biased or approximate method that avoids some MCMC moves may be useful if it provides a reasonable posterior approximation at substantially lower cost.

This motivates the ABC experiments in the next section. In ABC, each MCMC move typically requires simulation from the model, which can be the dominant computational cost. We therefore investigate whether Pareto smoothing can be used as a cheaper mechanism for mitigating degeneracy in ABC-SMC, where the aim
is not necessarily to improve on standard SMC at equal computational effort, but to trade a controlled loss of accuracy for a reduction in simulator calls. We use the adaptive algorithms to do this, where the number of MCMC moves is limited since they are not implemented until $\hat{k}>0.7$.

\section{Use in approximate Bayesian computation} \label{sec:abc}

\subsection{ABC, SMC and ABC-SMC\label{subsec:ABC,-SMC-and}}

We consider the challenge of estimating the posterior distribution
\(\pi(\theta\mid y_{\mathrm{obs}})\) of a parameter vector \(\theta\) given
observed data \(y_{\mathrm{obs}}\), in the case where the prior \(p(\theta)\)
is available analytically but the observation model \(f(y\mid\theta)\)
cannot be evaluated pointwise at \(\theta\). We assume, however, that
the model can be simulated from, \(y\sim f(\cdot\mid\theta)\). This challenge
is sometimes known as simulation-based inference (SBI), with approximate
Bayesian computation (ABC) being one of the most commonly used approaches to
approximating the posterior \(\pi\).

ABC makes use of an approximate likelihood, the \emph{ABC likelihood},
\begin{equation}
\int P_{\epsilon}\left(s_{\mathrm{obs}}\mid s\right)f_{S}\left(s\mid\theta\right)\,ds,
\label{eq:abc_llhd}
\end{equation}
where \(s=S(y)\) is a vector of summary statistics, with
\(s_{\mathrm{obs}}=S(y_{\mathrm{obs}})\), \(f_{S}\) is the distribution of
\(S(y)\) induced by \(f(y\mid\theta)\), and \(P_{\epsilon}\) is the
\emph{ABC kernel}. The kernel \(P_{\epsilon}\) takes values in
\([0,\infty)\) and is chosen to take higher values when \(s\) is closer
to \(s_{\mathrm{obs}}\), with the \emph{tolerance} \(\epsilon>0\)
controlling the ``width'' of the kernel. We refer to the posterior resulting
from using the ABC likelihood with tolerance \(\epsilon\) as the
\emph{ABC posterior} with tolerance \(\epsilon\). This likelihood is estimated
using \(P_{\epsilon}\left(s_{\mathrm{obs}}\mid s\right)\), where
\(s=S(y)\) and \(y\sim f(\cdot\mid\theta)\), corresponding to a Monte Carlo
estimate of equation~\eqref{eq:abc_llhd} using a single simulated dataset.
When the summary statistics are sufficient, the ABC likelihood tends to the
true likelihood \(f(y_{\mathrm{obs}}\mid\theta)\) as
\(\epsilon\rightarrow 0\).

This likelihood approximation is used in Monte Carlo methods for simulating
from an approximate posterior distribution on \(\theta\). SMC is particularly popular in the ABC literature, and ABC-SMC
methods were introduced and popularised in
\citet{sisson2007smcabc,toni2009abcsmc,delmoral2012adaptiveabc}. In ABC-SMC
algorithms, the sequence of targets is chosen to be
\begin{equation}
\pi_{t}\left(\theta,y\mid s_{\mathrm{obs}}\right)
\propto
p(\theta)f\left(y\mid\theta\right)
P_{\epsilon_{t}}\left(s_{\mathrm{obs}}\mid S(y)\right),
\end{equation}
where \(\infty=\epsilon_{0}>\cdots>\epsilon_{T}=\epsilon\). The tolerance
\(\epsilon\) is chosen to be small enough to provide an appropriate trade-off
between the accuracy of the corresponding ABC posterior and the computational
cost required to estimate this ABC posterior accurately. Throughout, we take
\(P_{\epsilon_{t}}\) to be a Gaussian kernel with diagonal covariance matrix
whose entries are \(\epsilon_{t}^{2}\) in every dimension.

\subsection{Expensive simulators}

Much of the Bayesian computation literature is focused on developing
Monte Carlo methods that are unbiased, or asymptotically unbiased,
and for which variance is reduced as more computational effort is expended.
This gives practitioners some assurance that they can reduce errors
resulting from the inference process by increasing computational
effort.

The rate-determining step in SBI is usually simulation from \(f\).
Rather than being able to increase computational effort indefinitely,
a practitioner usually faces a limit on the number of times simulation
can be performed: for example, a computing cluster may impose a limit
on the number of hours that a single job can run. The practitioner's
task is then to provide the most accurate inference possible
within this time limit, and they may be willing to accept a small amount
of bias in their inference as a trade-off for substantially reduced
variance. Such a trade-off is a key component of many SBI methods.

In ABC, the choice of tolerance governs the bias--variance trade-off
in the ABC posterior, with bias decreasing and variance increasing as
\(\epsilon\) decreases. Monte Carlo methods, such as ABC-SMC, then aim
to estimate the posterior without bias, or with minimal bias.

In this section, we use different configurations of
Algorithm~\ref{alg:generic_pssmc} on a range of problems. In the ABC case,
the MCMC move involves simulating from the model; therefore, this is the
rate-determining step. We investigate configurations of
algorithm~\ref{alg:generic_pssmc} that use PS, attempting to reduce the
number of MCMC moves required. By using PS, we introduce bias into the SMC, with the aim of reducing variance.

\subsection{Related work}

The effect of PS is that points away from the high-weight regions gain
proportionally more weight, spreading out the particle population more than
would otherwise be the case. In this respect, the net effect of PS is not
dissimilar to ``jittering'' \citep{gordon1993novel,crisan2018nested} the
particle population; that is, adding Gaussian noise with covariance
\(\Sigma_t/N^{3/2}\). Here, \(\Sigma_t\) is chosen to give larger scaling to
dimensions with larger support, and the term \(1/N^{3/2}\) guarantees control
of the error introduced by jittering.

The use of Gaussian kernels to move particles in SMC is routine, and is usually
incorporated into a Metropolis--Hastings move and paired with an appropriate
reweighting step. The use of Gaussian kernels without an accept--reject step is
also common, resulting in marginal SMC or population Monte Carlo schemes that
also have an appropriate reweighting step.

Like PS, jittering introduces bias, but with the benefit of reducing the number
of MCMC moves required in SMC, and therefore reducing computational cost. To
our knowledge, jittering has not been used previously in the ABC context, and
we compare it with PS in our experiments below. In all experiments, after
resampling, we jitter each particle according to
\[
\theta_t^i = \theta_t^{i*} + \eta_t^i,
\qquad
\eta_t^i \sim N\!\left(0,\frac{\Sigma_t}{N^{3/2}}\right),
\]
where \(\Sigma_t = c^2 \widehat{\mathrm{Var}}(\theta_t^{1:N})\), with
\(\widehat{\mathrm{Var}}(\theta_t^{1:N})\) denoting the sample covariance
matrix of the current particle population. No Metropolis--Hastings
accept--reject step or additional importance reweighting is applied after
jittering, so this can be viewed as a biased heuristic move used as a
cheaper alternative to the usual MCMC move.

We also compare the PS approaches with an approach that truncates the weights
at each iteration, as in equation~\eqref{eq:truncation}, which we call T-SMC.

\subsection{Empirical studies}

The details of the experiments are the same in each subsection unless stated.
We used \(N=1000\) particles per ABC-SMC run. A pilot run was first used to
construct an adaptive tolerance schedule \(\{\varepsilon_t\}_{t=1}^T\), chosen
so that at each step the effective sample size was \(99\%\) of that under the
previous tolerance. This tolerance schedule was then fixed and used for all
methods in the comparison. Standard SMC performs a resample--move step at every
tolerance level. The adaptive methods a-SMC, aPS-SMC, and aPS-SIS/SMC use the
same sequence of tolerances, but only perform MCMC moves when \(\hat{k}\)
exceeds 0.7, and as such typically have a lower computational cost than SMC. The T-SMC and jitter methods also use this same tolerance schedule; for the
jitter methods we report results for \(c=1,0.1,0.01\). As a reference, we also perform an ABC-MCMC run of \(10^7\) iterations to approximate the ``true'' posterior. All proposals in both SMC and MCMC are Gaussian random walks in parameter space.

\subsubsection{\texorpdfstring{$g$-and-$k$}{g-and-k} distribution}
\label{sec:gandk}

In this section, we study the four-parameter g-and-k distribution to the daily US/Canadian dollar exchange-rate log-returns from the \texttt{Garch} dataset in the \texttt{Ecdat} R package \citep{rayner2002gk,drovandi2011indirect}. Denote by \(x_t = \log\bigl(r_{t+1}/r_t\bigr)\) the log-return on day \(t\), where \(r_t\) is the spot rate. The g-and-k distribution is defined via its quantile function
\[
Q(p;A,B,g,k,c)\;=\;A \;+\; B\Bigl(1 + c\,\tanh\!\bigl(\tfrac{g\,z}{2}\bigr)\Bigr)\,z\,(1+z^2)^k,
\quad z=\Phi^{-1}(p),
\]
with location \(A\), scale \(B>0\), skewness \(g\), tail parameter \(k\ge 0\), and fixed \(c=0.8\) \citep{rayner2002gk}. We draw trajectories of IID g-and-k variates via the \texttt{rgk()} function in the R package \texttt{gk} \citep{prangle2020gk}.

The observed data \(\{x_t\}_{t=1}^N\) consist of \(N=1966\) log-returns (1980--1987). We reduce these to a 4-dimensional summary
\begin{align}
    S_A &= E_{(N/2)}, \\
    S_B &= E_{(3N/4)} - E_{(N/4)}, \\
    S_g &= \frac{E_{(3N/4)} + E_{(N/4)} - 2\,E_{(N/2)}}{S_B}, \\
    S_k &= \frac{E_{(N)} - E_{(3N/4)} + E_{(N/2)} - E_{(N/4)}}{S_B},
\end{align}
where \(E_{(i)}\) is the \(i\)th order statistic of the sample and, for efficiency, is approximated by simulating just the seven octiles via the \texttt{orderstats()} function \citep{fearnhead2012semiauto,prangle2020gk,ripley1987stochastic}.

Uniform priors are placed on
\[
A\sim U(-1,1),\quad B\sim U(0,1),\quad g\sim U(-5,5),\quad k\sim U(0,10).
\]

\begin{table}
\caption{Wasserstein distances and computational performance for ABC-SMC variants on the g-and-k experiment. The Wasserstein columns report distances between the ABC-SMC posteriors and the reference posterior for each parameter. For each column, the best (smallest) value among the non-Standard variants is shown in \textbf{bold}.}
\label{tab:abc_smc_wasserstein_performance_gk}
\small
\begin{tabular*}{\tblwidth}{@{\extracolsep{\fill}}lcccccc@{}}
\toprule
ABC-SMC Method
& \(A\)
& \(B\)
& \(g\)
& \(k\)
& \begin{tabular}[c]{@{}c@{}}Avg.\ runtime\\(s)\end{tabular}
& \begin{tabular}[c]{@{}c@{}}Avg.\ \# resampling\\steps\end{tabular} \\
\midrule
SMC
& \(6.01\times10^{-4}\)
& \(4.59\times10^{-3}\)
& \(2.02\times10^{-2}\)
& \(3.56\times10^{-2}\)
& 57.98
& 115.32 \\
a-SMC
& \(1.03\times10^{-2}\)
& \(1.02\times10^{-1}\)
& \(1.38\times10^{-1}\)
& \(4.54\times10^{-1}\)
& 14.50
& 3.61 \\
aPS-SMC
& \(9.70\times10^{-3}\)
& \(8.62\times10^{-2}\)
& \(1.24\times10^{-1}\)
& \(4.21\times10^{-1}\)
& 13.76
& 3.56 \\
aPS-SIS/SMC
& \(7.35\times10^{-3}\)
& \(5.93\times10^{-2}\)
& \(1.05\times10^{-1}\)
& \(3.30\times10^{-1}\)
& \textbf{12.89}
& \textbf{3.31} \\
T-SMC
& \(\mathbf{5.21\times10^{-3}}\)
& \(5.88\times10^{-2}\)
& \(1.15\times10^{-1}\)
& \(4.11\times10^{-1}\)
& 13.40
& 7.62 \\
Jitter, \(c=1\)
& \(1.19\times10^{-1}\)
& \(2.68\times10^{-1}\)
& \(6.37\times10^{-1}\)
& \(1.73\times10^{0}\)
& 38.72
& 14.82 \\
Jitter, \(c=0.1\)
& \(5.99\times10^{-3}\)
& \(\mathbf{5.76\times10^{-2}}\)
& \(\mathbf{7.64\times10^{-2}}\)
& \(\mathbf{2.65\times10^{-1}}\)
& 26.09
& 4.25 \\
Jitter, \(c=0.01\)
& \(6.57\times10^{-3}\)
& \(7.13\times10^{-2}\)
& \(9.32\times10^{-2}\)
& \(3.28\times10^{-1}\)
& 26.11
& 3.63 \\
\bottomrule
\end{tabular*}
\end{table}

\begin{figure}
  \centering
  \includegraphics[width=\textwidth]{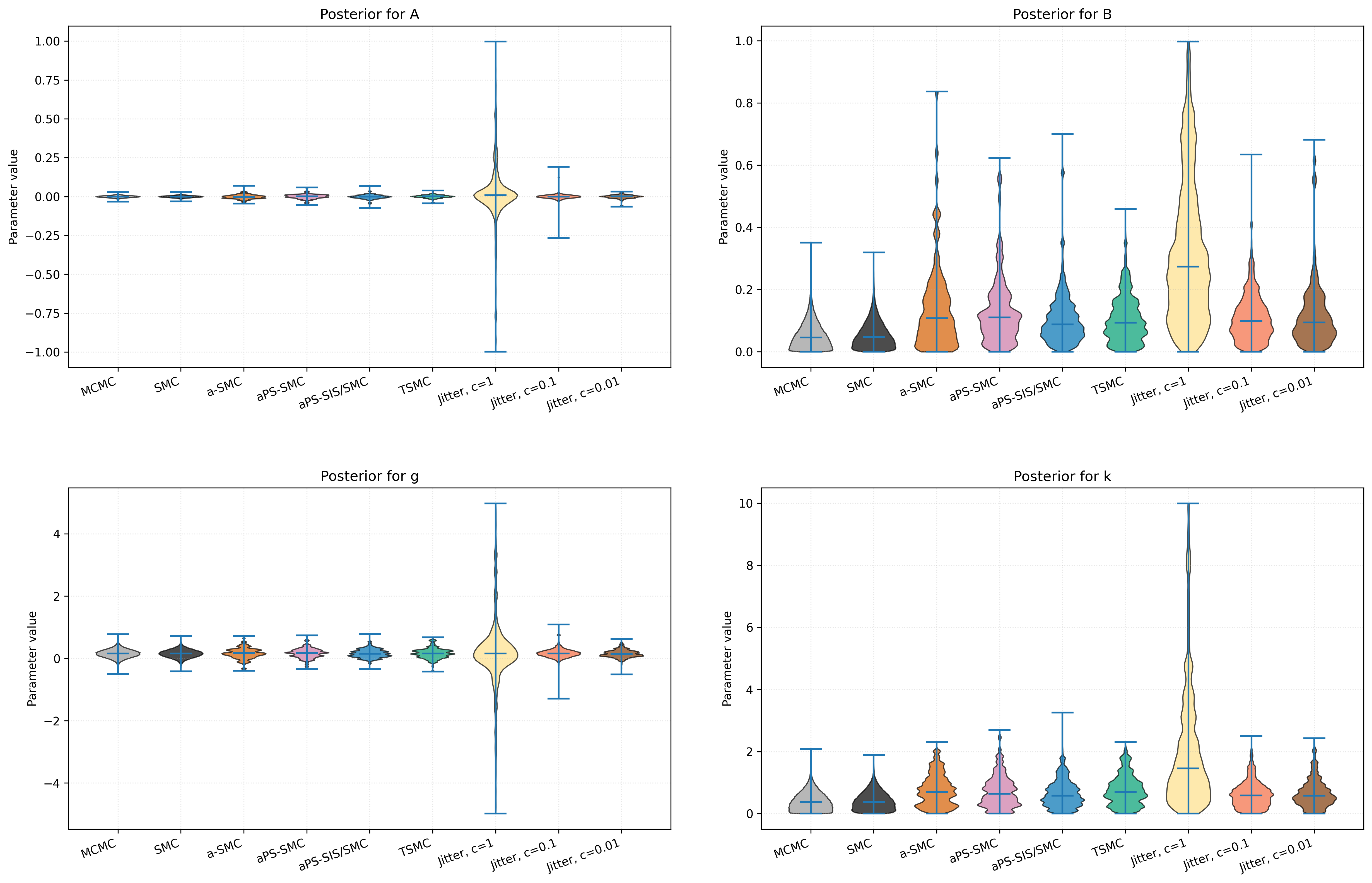}
  \caption{g–and–k model: pooled posterior violin plots across ABC-SMC variants.
  Each panel shows the empirical posterior distributions aggregated over 1{,}000 independent runs for a single parameter (\(A,B,g,k\)).
  Violin width encodes density, and the y-axis is the parameter value. Methods appear in the same left-to-right order across panels.}
  \label{fig:gk_violin_all}
\end{figure}

\subsubsection{Lotka--Volterra model}
\label{sec:lv}

In this section, we apply our ABC-SMC methodologies to the stochastic Lotka--Volterra (LV) predator--prey model, cast as a three-parameter Markov jump process \citep{wilkinson2013summary,papamakarios2016fast}. Denote by \(x_1\) and \(x_2\) the predator and prey populations, respectively. The system evolves according to:
\begin{itemize}
  \item \textbf{Prey birth:}
    \(\;x_2 \to x_2 + 1\) at rate \(k_1\,x_2\).
  \item \textbf{Predation:}
    \((x_1, x_2) \to (x_1 + 1, x_2 - 1)\) at rate \(k_2\,x_1\,x_2\).
  \item \textbf{Predator death:}
    \(\;x_1 \to x_1 - 1\) at rate \(k_3\,x_2\).
\end{itemize}
Let \(x_{t,1}\) and \(x_{t,2}\) denote predator and prey counts at time \(t\), respectively, and write \(\mathbf{x}_t = (x_{t,1},x_{t,2})^\top\). We generate trajectories using the Gillespie algorithm in Python, adapted from the \texttt{smfsb} R package \citep{wilkinson2024smfsb}. The observed data \(\mathbf{y}_{\mathrm{obs}}\) comprise the \texttt{LVperfect} series—an initial state at \(t=0\) plus 15 subsequent bi-population measurements. We then compute summary statistics \(s_{\mathrm{obs}}\) as in \citet{wilkinson2013summary}, reducing the data to a nine-dimensional vector. Independent log-uniform priors are placed on the three rate parameters:
\[
\log k_1 \sim U(-6,2),\qquad
\log k_2 \sim U(-6,2),\qquad
\log k_3 \sim U(-6,2),
\]
equivalently,
\[
k_1,k_2,k_3 \sim \mathrm{LogUniform}(e^{-6},e^2).
\]

\begin{table}
\caption{Wasserstein distances and computational performance for ABC-SMC variants on the Lotka--Volterra experiment. The Wasserstein columns report distances between the ABC-SMC posteriors and the reference posterior for each parameter. For each column, the best (smallest) value among the non-Standard variants is shown in \textbf{bold}.}
\label{tab:abc_smc_wasserstein_performance_lv}
\small
\begin{tabular*}{\tblwidth}{@{\extracolsep{\fill}}lccccc@{}}
\toprule
Method
& \(k_1\)
& \(k_2\)
& \(k_3\)
& \begin{tabular}[c]{@{}c@{}}Runtime\\(s)\end{tabular}
& \begin{tabular}[c]{@{}c@{}}\# Resamples\\per run\end{tabular} \\
\midrule
SMC
& \(4.06\times10^{-2}\)
& \(1.88\times10^{-4}\)
& \(2.75\times10^{-2}\)
& 3439.76
& 96.42 \\
a-SMC
& \(2.51\times10^{-1}\)
& \(1.59\times10^{-3}\)
& \(2.44\times10^{-1}\)
& 334.40
& 9.32 \\
aPS-SMC
& \(2.92\times10^{-1}\)
& \(1.45\times10^{-3}\)
& \(2.33\times10^{-1}\)
& 817.92
& 10.79 \\
aPS-SIS/SMC
& \(2.15\times10^{-1}\)
& \(1.32\times10^{-3}\)
& \(2.26\times10^{-1}\)
& \textbf{319.85}
& 9.90 \\
T-SMC
& \(\mathbf{1.82\times10^{-1}}\)
& \(\mathbf{8.30\times10^{-4}}\)
& \(\mathbf{1.51\times10^{-1}}\)
& 319.95
& \textbf{9.12} \\
Jitter, \(c=1\)
& \(3.44\times10^{0}\)
& \(7.62\times10^{-1}\)
& \(3.31\times10^{0}\)
& 800.94
& 31.11 \\
Jitter, \(c=0.1\)
& \(3.15\times10^{0}\)
& \(3.87\times10^{-3}\)
& \(3.10\times10^{0}\)
& 763.80
& 24.53 \\
Jitter, \(c=0.01\)
& \(2.57\times10^{0}\)
& \(1.64\times10^{-3}\)
& \(2.43\times10^{0}\)
& 492.05
& 12.09 \\
\bottomrule
\end{tabular*}
\end{table}

\begin{figure}
  \centering
  \includegraphics[width=\textwidth]{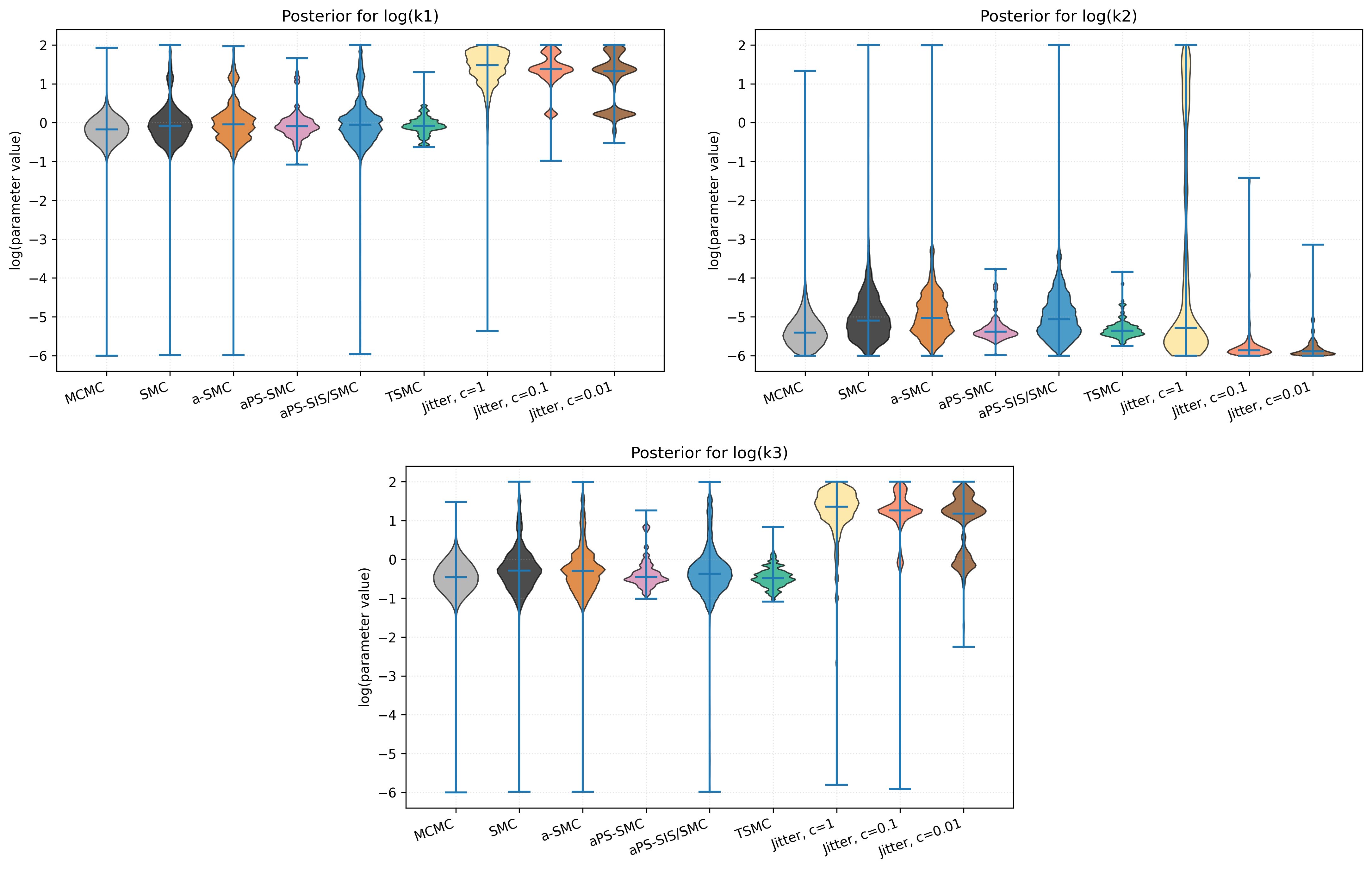}
  \caption{Lotka--Volterra model: pooled posterior violin plots across ABC-SMC variants. Each violin summarises the empirical posterior draws aggregated over 1{,}000 independent runs for the indicated parameter; violin width reflects density. The methods are shown in the same left-to-right order as in the plots (MCMC, SMC, a-SMC, aPS-SMC, aPS-SIS/SMC, T-SMC, Jitter, \(c=1\), Jitter, \(c=0.1\), and Jitter, \(c=0.01\)).}
  \label{fig:lv_violin_all}
\end{figure}

\subsubsection{Ricker model}
\label{sec:ricker}

In this section, we apply our ABC-SMC methodologies to the stochastic Ricker population model, a simple but challenging one-dimensional time series exhibiting over-compensation \citep{ricker1954stock,wood2010statistical}. Denote by \(S_t\) the true (latent) population at discrete time \(t\) and by \(y_t\) the noisy count. The dynamics are
\[
\log S_t = \log r + \log S_{t-1} - S_{t-1} + \sigma\,\epsilon_t,
\quad \epsilon_t\sim N(0,1),
\]
\[
y_t \mid S_t \;\sim\; \mathrm{Poisson}\bigl(\psi\,S_t\bigr).
\]
We generate a synthetic ``observed'' data set \(\mathbf{y}_{\mathrm{obs}} = (y_1,\dots,y_T)^\top\) of length \(T=50\) under fixed parameters \((\log r,\log \sigma,\log \psi)\) chosen within the prior support. Summary statistics \(s_{\mathrm{obs}}\in\mathbb{R}^{13}\) are computed following Wood's thirteen-dimensional set of autocovariances, regression coefficients, and moment-based features \citep{wood2010statistical}.

Priors on the log-parameters are
\[
\log r      \sim U(2,5),\quad
\log \sigma \sim U(-3.0,-2.3),\quad
\log \psi   \sim U(-1.79,1.61).
\]

\begin{table}
\caption{Wasserstein distances and computational performance for ABC-SMC variants on the Ricker experiment. The Wasserstein columns report distances between the ABC-SMC posteriors and the reference posterior for each parameter. For each column, the best (smallest) value among the non-Standard variants is shown in \textbf{bold}.}
\label{tab:abc_smc_wasserstein_performance_ricker}
\small
\begin{tabular*}{\tblwidth}{@{\extracolsep{\fill}}lccccc@{}}
\toprule
Method
& \(\log R\)
& \(\log \sigma\)
& \(\log \phi\)
& \begin{tabular}[c]{@{}c@{}}Runtime\\(s)\end{tabular}
& \begin{tabular}[c]{@{}c@{}}\# Resamples\\per run\end{tabular} \\
\midrule
SMC
& \(1.20\times10^{-1}\)
& \(1.25\times10^{-1}\)
& \(5.45\times10^{-2}\)
& 167.88
& 83.01 \\
a-SMC
& \(1.67\times10^{-1}\)
& \(1.84\times10^{-1}\)
& \(8.93\times10^{-2}\)
& 19.79
& 3.18 \\
aPS-SMC
& \(1.39\times10^{-1}\)
& \(1.68\times10^{-1}\)
& \(1.25\times10^{-1}\)
& \textbf{11.44}
& 2.59 \\
aPS-SIS/SMC
& \(1.55\times10^{-1}\)
& \(1.42\times10^{-1}\)
& \(8.43\times10^{-2}\)
& 19.01
& 3.60 \\
T-SMC
& \(1.32\times10^{-1}\)
& \(1.45\times10^{-1}\)
& \(7.19\times10^{-2}\)
& 13.18
& \textbf{2.24} \\
Jitter, \(c=1\)
& \(3.61\times10^{-1}\)
& \(\mathbf{5.99\times10^{-2}}\)
& \(3.09\times10^{-1}\)
& 25.46
& 7.92 \\
Jitter, \(c=0.1\)
& \(\mathbf{8.01\times10^{-2}}\)
& \(1.01\times10^{-1}\)
& \(\mathbf{5.68\times10^{-2}}\)
& 19.65
& 3.81 \\
Jitter, \(c=0.01\)
& \(1.15\times10^{-1}\)
& \(1.19\times10^{-1}\)
& \(6.37\times10^{-2}\)
& 48.61
& 4.62 \\
\bottomrule
\end{tabular*}
\end{table}

\begin{figure}
  \centering
  \includegraphics[width=\textwidth]{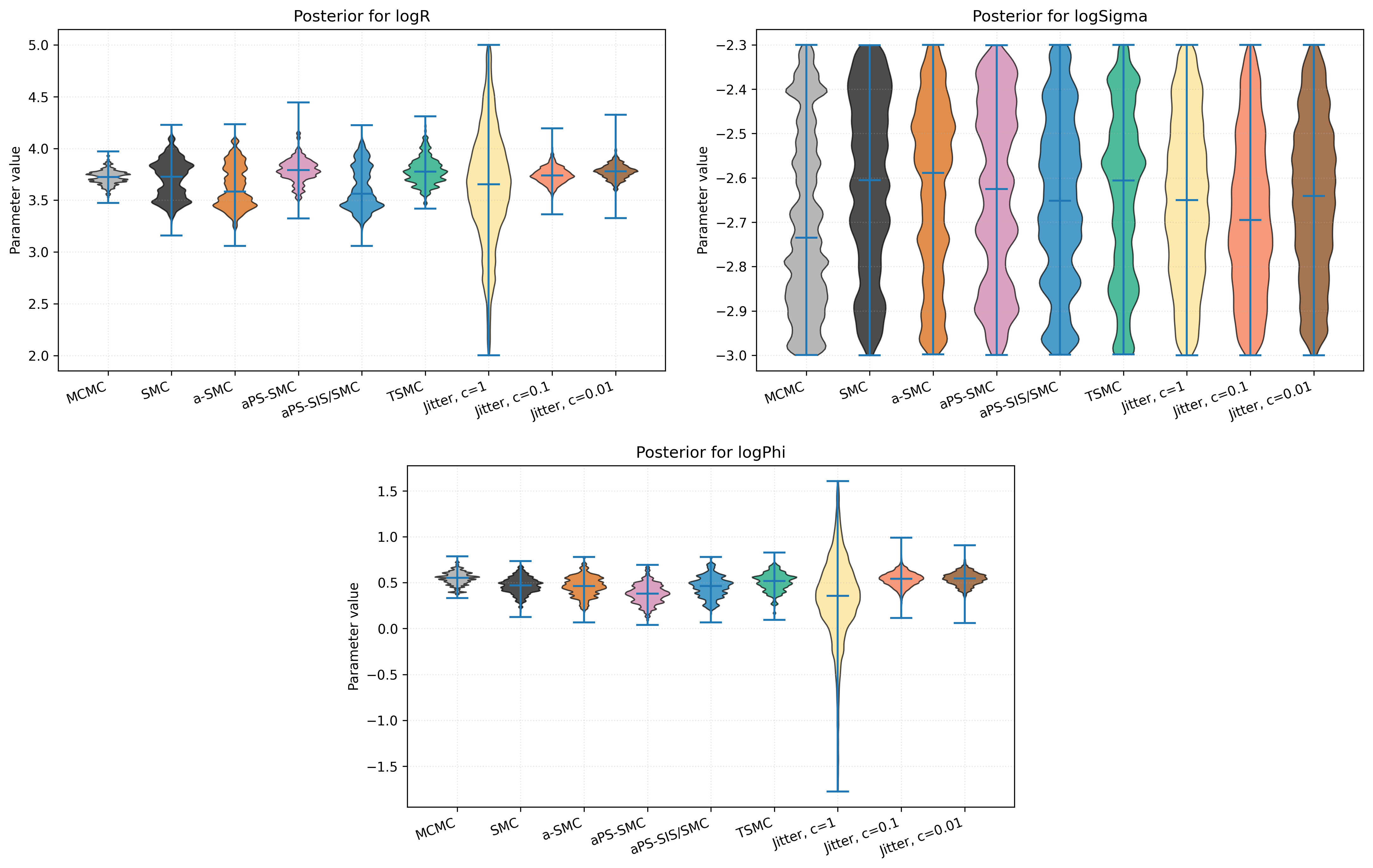}
  \caption{Ricker model: pooled posterior violin plots across ABC-SMC variants.
  Each violin summarises the empirical posterior draws aggregated over 1{,}000 independent runs for the indicated parameter; violin width reflects density. The methods are shown in the same left-to-right order as in the plots (MCMC, SMC, a-SMC, aPS-SMC, aPS-SIS/SMC, T-SMC, Jitter, \(c=1\), Jitter, \(c=0.1\), and Jitter, \(c=0.01\)).}
  \label{fig:ricker_violin_all}
\end{figure}

\subsection{Review of results}

Across the three ABC examples, no single reduced-cost approximation strategy
uniformly outperforms the others. Standard ABC-SMC gives the closest agreement
with the reference ABC posterior in all three examples, as expected, but at a
substantially higher computational cost because it performs many more
resample--move steps. The relevant question is therefore whether the
reduced-cost methods can provide useful computational savings while retaining
an acceptable approximation to the reference posterior.


The experiments do not provide clear evidence that PS is
preferable to simpler weight truncation. T-SMC is often as accurate as, and sometimes more accurate than, the Pareto-based alternatives. This is especially clear in the Lotka--Volterra experiment, where T-SMC gives the smallest
Wasserstein distances among the reduced-cost methods for all three parameters, while also requiring the fewest average resampling steps.

The jittering methods are more sensitive to the tuning parameter \(c\). Large
jitter, corresponding to \(c=1\), is generally unreliable, whereas smaller
amounts of jitter can be competitive. In particular, jittering with \(c=0.1\)
performs well in the g-and-k and Ricker examples, including giving the smallest
reduced-cost Wasserstein distances for several parameters. This indicates that
jittering can be effective when the perturbation scale is well matched to the
problem, but its performance is less robust to tuning than the adaptive
weight-based methods.

Overall, standard ABC-SMC remains the most reliable method in terms of posterior accuracy, but at a substantially larger computational cost. Among the reduced-cost alternatives, \(\hat{k}\)-adaptive methods, truncation, and
carefully tuned jittering can all reduce runtime substantially, but their accuracy is problem-dependent.


\section{Discussion}


This paper investigated whether Pareto smoothing can improve SMC and ABC-SMC algorithms, either by stabilising importance weights or by reducing the number of expensive resample--move steps. The motivation is that SMC samplers are built
from repeated importance sampling updates, while PSIS provides both a regularisation of large weights and the diagnostic \(\hat{k}\), which measures the tail behaviour of the weight distribution.

The main finding is that, whilst we find some evidence that the weight adjustment in PS can lead to improved estimators, it is not likely to have a significant impact when used in SMC algorithms in most cases. The reason is that in settings where PS might be used to reduce variance when performing IS between two distributions, the SMC-based alternative solution would be to introduce an intermediate distribution and use SMC instead. Sequences of distributions typically used in SMC are not those where PS shows large benefits.

One concern of using PS at every step of an SMC is that it introduces bias that could grow unboundedly. When used as an additional step at each iteration of an otherwise-standard SMC sampler, we do not see evidence of this issue. However, without MCMC moves (in the PS-SIS results), we do observe this. These findings are consistent with the ideas in \cite{everitt2017bayesian} about bias in weight updates in SMC.

Across a range of experiments, the setting where we found the clearest benefits for PS was in the ``variance expansion'' case, which is most reminiscent of the cross-validation scenarios where PS has been used extensively.

\section*{Acknowledgements}
Tan is supported by a CENTA (The Central England NERC Training Alliance) PhD studentship with CASE sponsorship from Cefas.

\clearpage
\appendix

\section{Additional results}

This appendix reports additional results for the ESS--\(\hat{k}\) experiments,
the Gaussian SMC sampler experiments, and the ABC-SMC experiments. For the
ESS--\(\hat{k}\) study, we provide further results beyond the one-dimensional
Gaussian examples shown in the main text, including the corresponding
\(100\)-dimensional Gaussian cases and additional exponential examples.

For the Gaussian SMC sampler examples, we give a more detailed decomposition of
the moment-estimation errors than is shown in the main text, separating the
mean squared error into squared bias and Monte Carlo variance for both the
first and second moments.

For the ABC-SMC experiments, we report additional repeated-run summaries. For
each method and parameter, we compute the posterior mean and posterior standard
deviation in each independent run and compare these quantities with the
corresponding values from the reference ABC-SMC posterior. The reported mean
squared error is again decomposed into variance and squared bias across
repeated runs. As in the main text, bold entries denote the best-performing
non-standard method in each column.

\subsection{Additional ESS--\texorpdfstring{\(\hat{k}\)}{k-hat} results}
\label{app:ess_khat_extra}

This appendix provides additional ESS--\(\hat{k}\) results supplementing
Section~\ref{sec:ESSkhat}. The main text considered three one-dimensional
Gaussian examples. Here we report analogous results for the corresponding
\(100\)-dimensional Gaussian examples, together with two exponential examples
closer to the settings used in \citet{vehtari2024psis}. The details of the proposal and target in each case are found in the captions of the figures. The method used to produce the figures is the same as in the main text.


\begin{figure}
    \centering

    \begin{subfigure}[t]{0.32\textwidth}
        \centering
        \includegraphics[width=\linewidth]{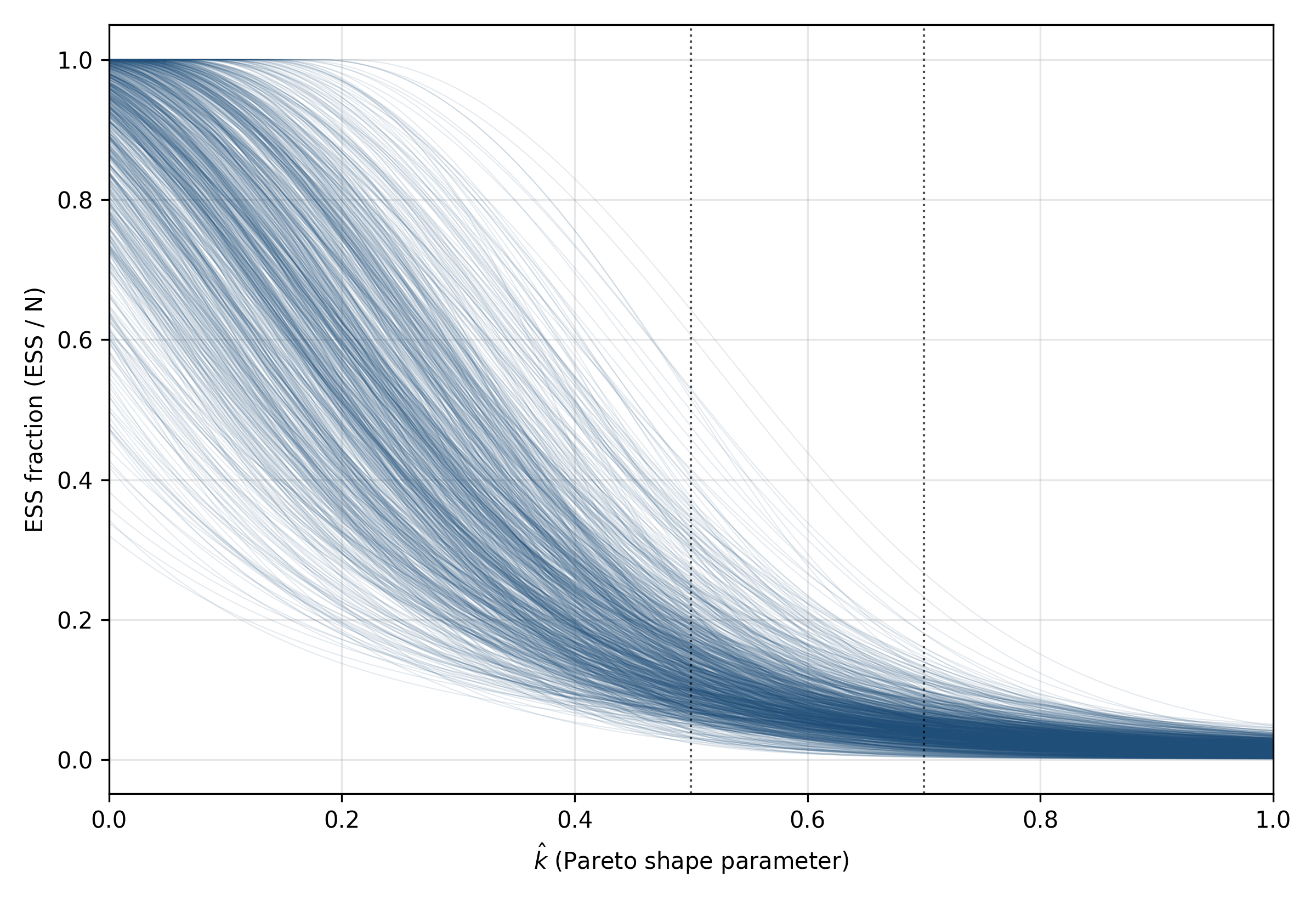}
        \caption{\(N(0,I_{100})\rightarrow N(10\mathbf{1},I_{100})\).}
        \label{fig:ess_khat_gaussian_d100_mean_shift}
    \end{subfigure}
    \hfill
    \begin{subfigure}[t]{0.32\textwidth}
        \centering
        \includegraphics[width=\linewidth]{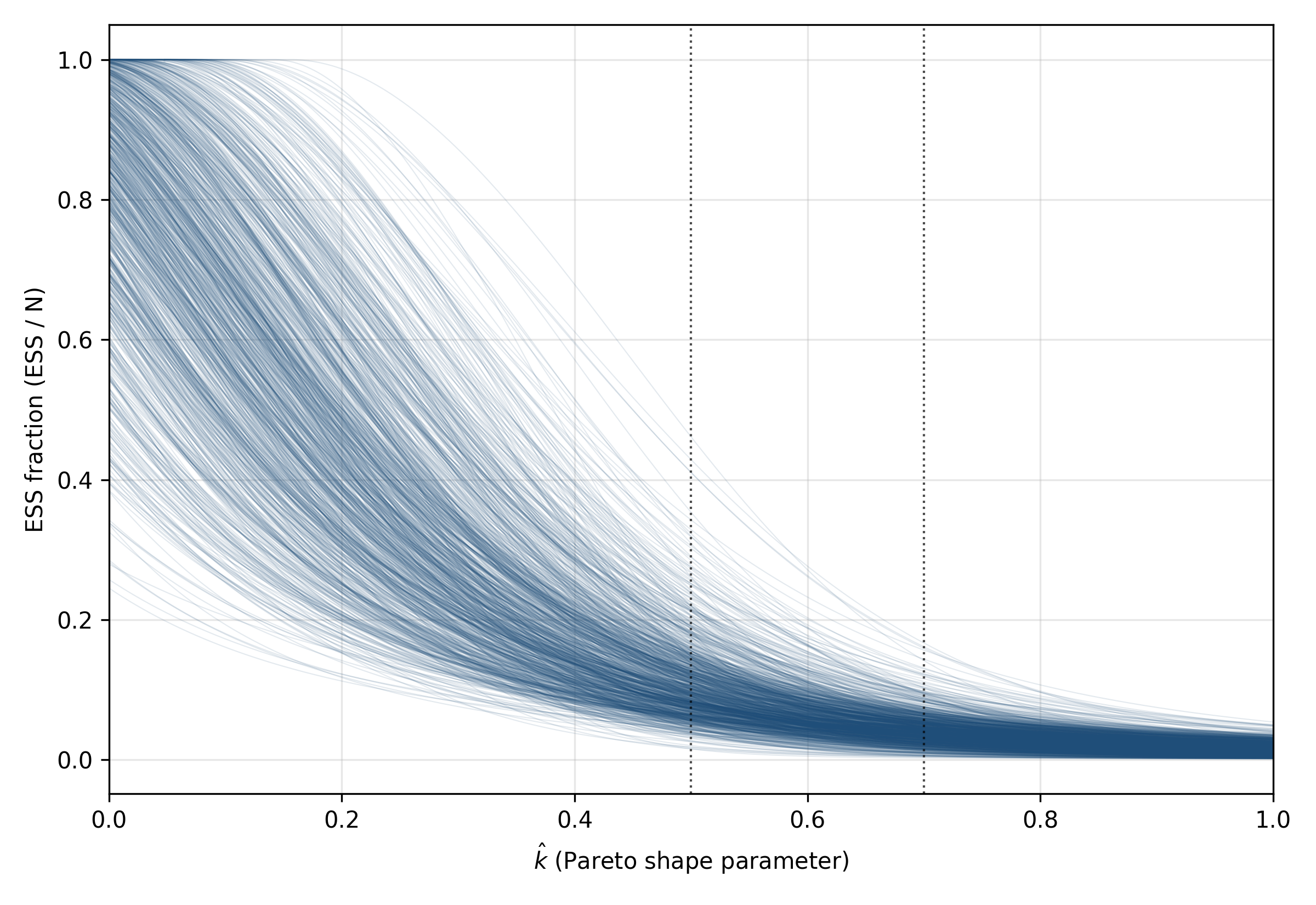}
        \caption{\(N(0,I_{100})\rightarrow N(0,10^{-5}I_{100})\).}
        \label{fig:ess_khat_gaussian_d100_thin}
    \end{subfigure}
    \hfill
    \begin{subfigure}[t]{0.32\textwidth}
        \centering
        \includegraphics[width=\linewidth]{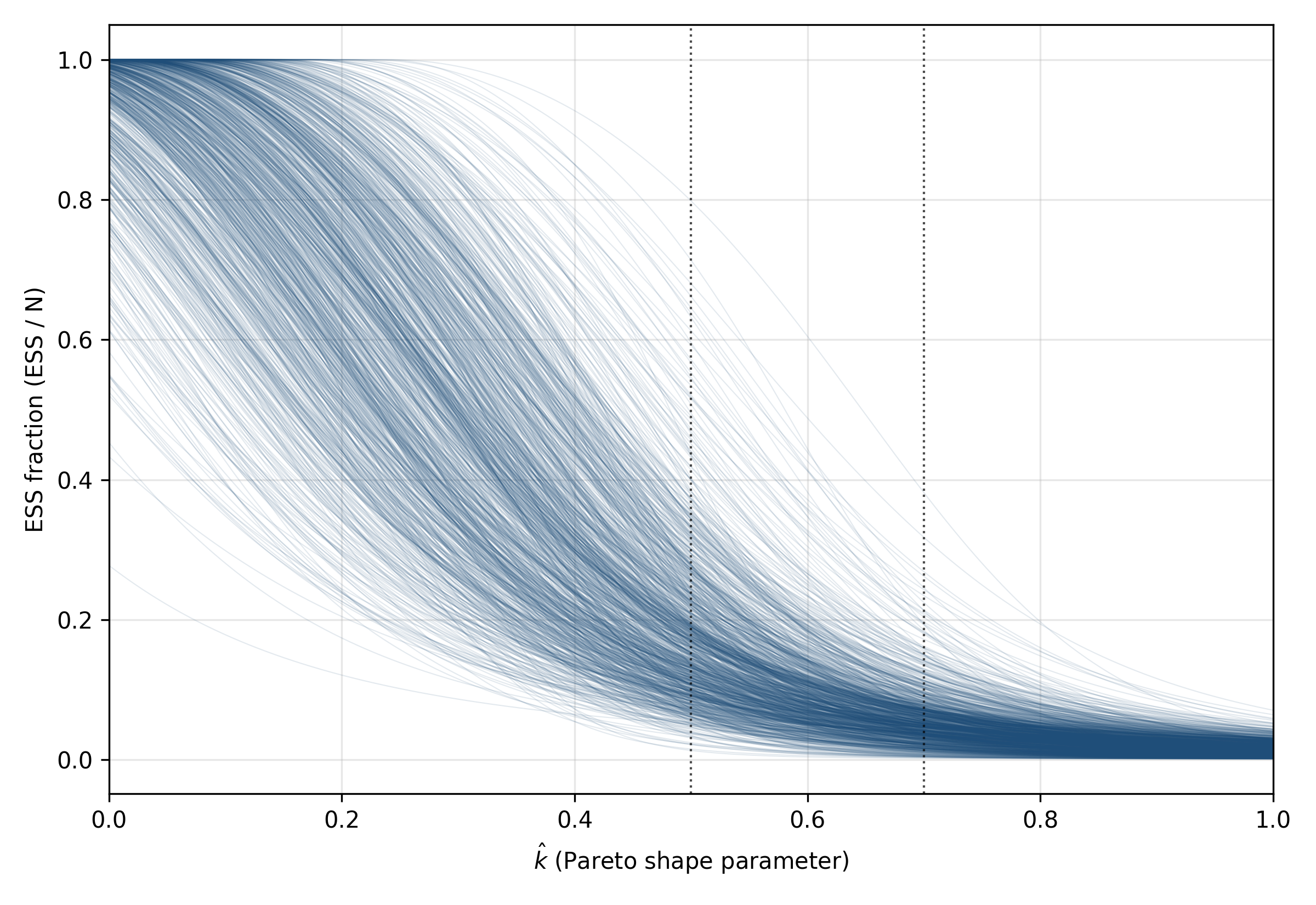}
        \caption{\(N(0,I_{100})\rightarrow N(0,10I_{100})\).}
        \label{fig:ess_khat_gaussian_d100_fat}
    \end{subfigure}

    \caption{ESS fraction \(\mathrm{ESS}/N\) versus \(\hat{k}\) for the
    three \(100\)-dimensional Gaussian examples. The qualitative behaviour is
    similar to the one-dimensional examples in Figure~\ref{fig:ess-v-khat}:
    larger \(\hat{k}\) generally corresponds to smaller ESS, but the precise
    relationship depends on the form of proposal--target mismatch.}
    \label{fig:ess_khat_gaussian_d100}
\end{figure}


\begin{figure}
    \centering

    \begin{subfigure}[t]{0.48\textwidth}
        \centering
        \includegraphics[width=\linewidth]{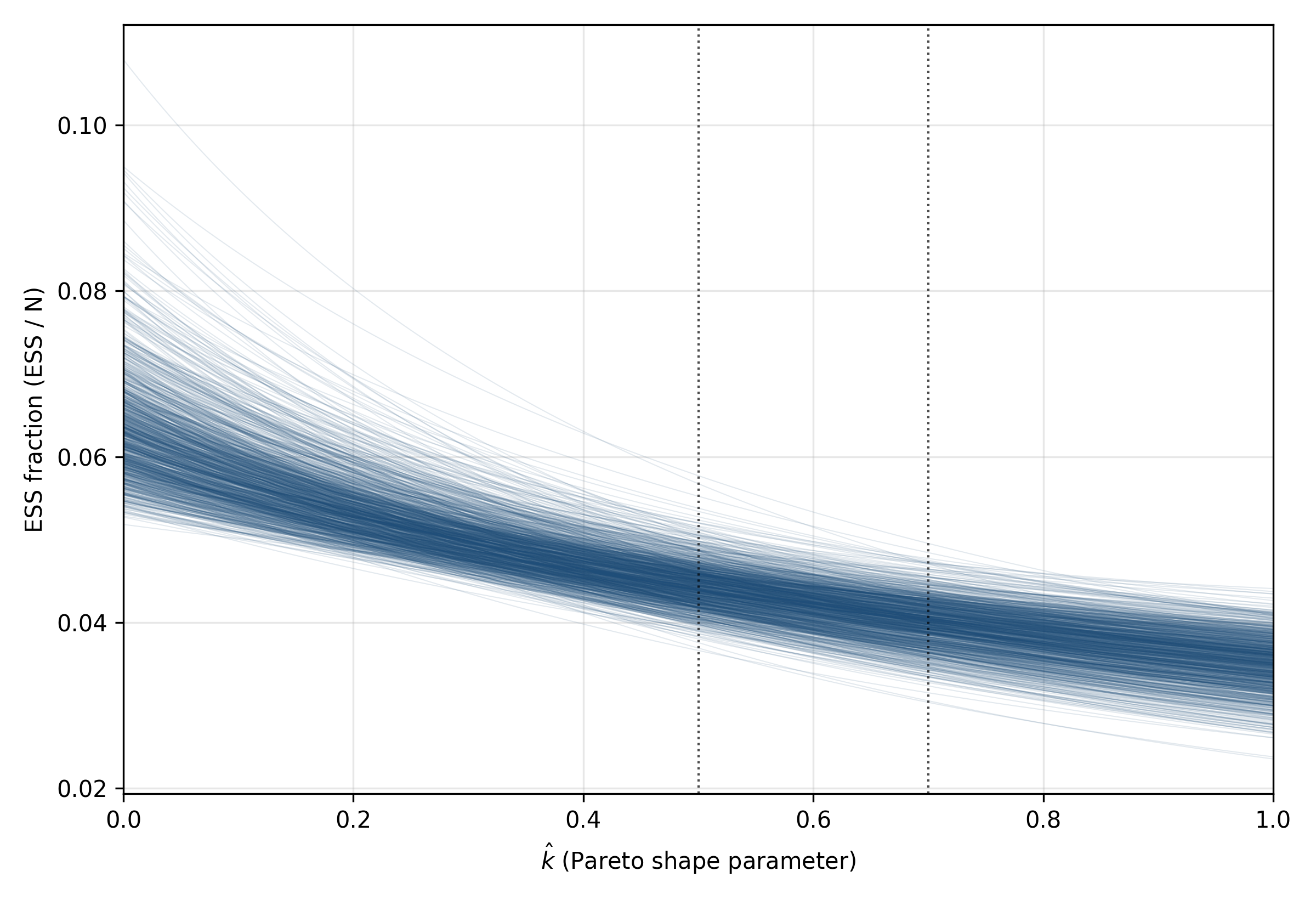}
        \caption{\(\mathrm{Exp}(1)\rightarrow\mathrm{Exp}(100)\).}
        \label{fig:ess_khat_exp_fat_to_thin}
    \end{subfigure}
    \hfill
    \begin{subfigure}[t]{0.48\textwidth}
        \centering
        \includegraphics[width=\linewidth]{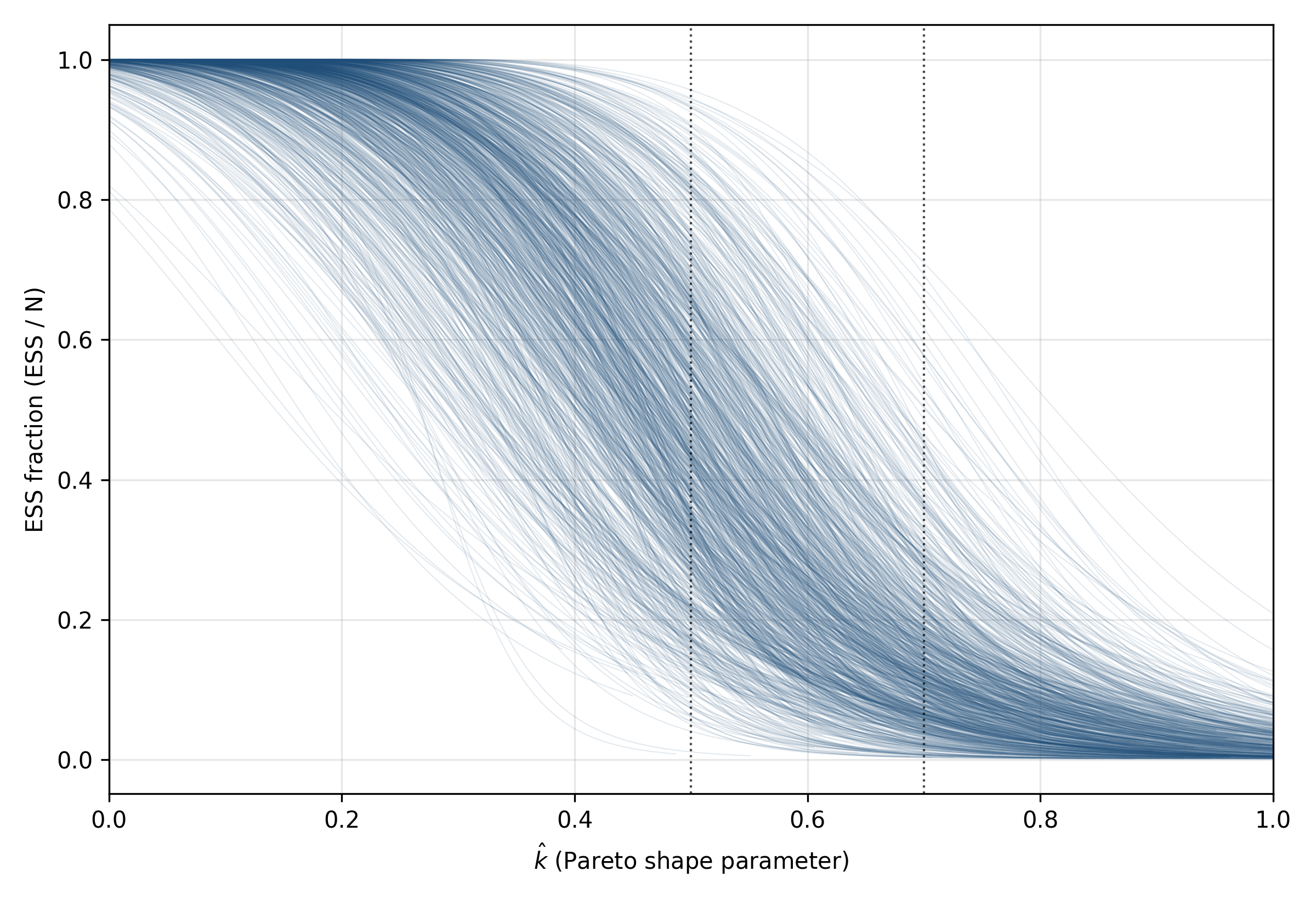}
        \caption{\(\mathrm{Exp}(1)\rightarrow\mathrm{Exp}(0.01)\).}
        \label{fig:ess_khat_exp_thin_to_fat}
    \end{subfigure}

    \caption{ESS fraction \(\mathrm{ESS}/N\) versus \(\hat{k}\) for two
    exponential examples. The proposal distribution is \(\mathrm{Exp}(1)\).
    The targets are \(\mathrm{Exp}(100)\), corresponding to a strong
    contraction, and \(\mathrm{Exp}(0.01)\), corresponding to a strong
    expansion.}
    \label{fig:ess_khat_exponential}
\end{figure}


\subsection{Additional Gaussian SMC sampler results}
\label{app:gaussian_smc_extra}

\begin{figure}
    \centering
    \includegraphics[width=\textwidth]{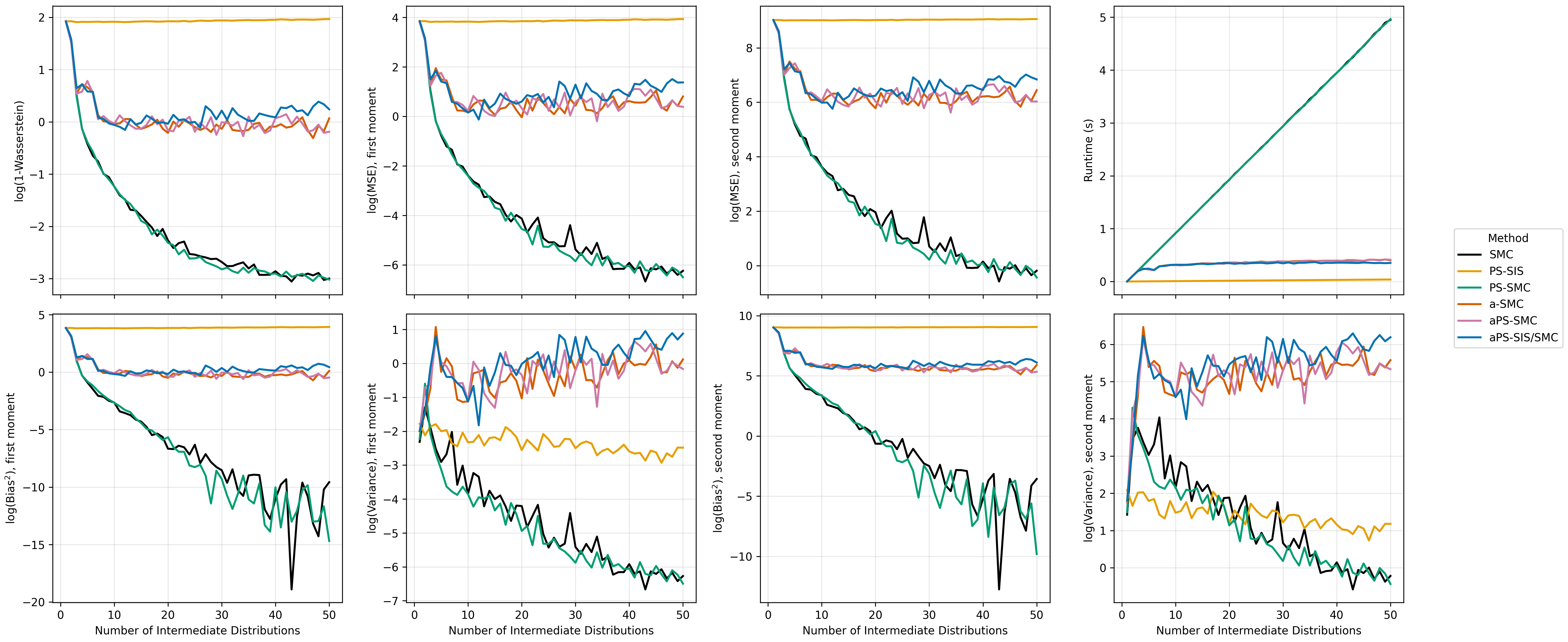}
    \caption{Detailed performance summaries for the mean-shift Gaussian SMC
    example \(N(0,1)\rightarrow N(10,1)\). The first row shows log
    Wasserstein distance, first-moment log MSE, second-moment log MSE, and
    runtime. The second row shows first-moment log squared bias,
    first-moment log variance, second-moment log squared bias, and
    second-moment log variance.}
    \label{fig:gaussian_smc_results_appendix_mean_shift}
\end{figure}

\begin{figure}
    \centering
    \includegraphics[width=\textwidth]{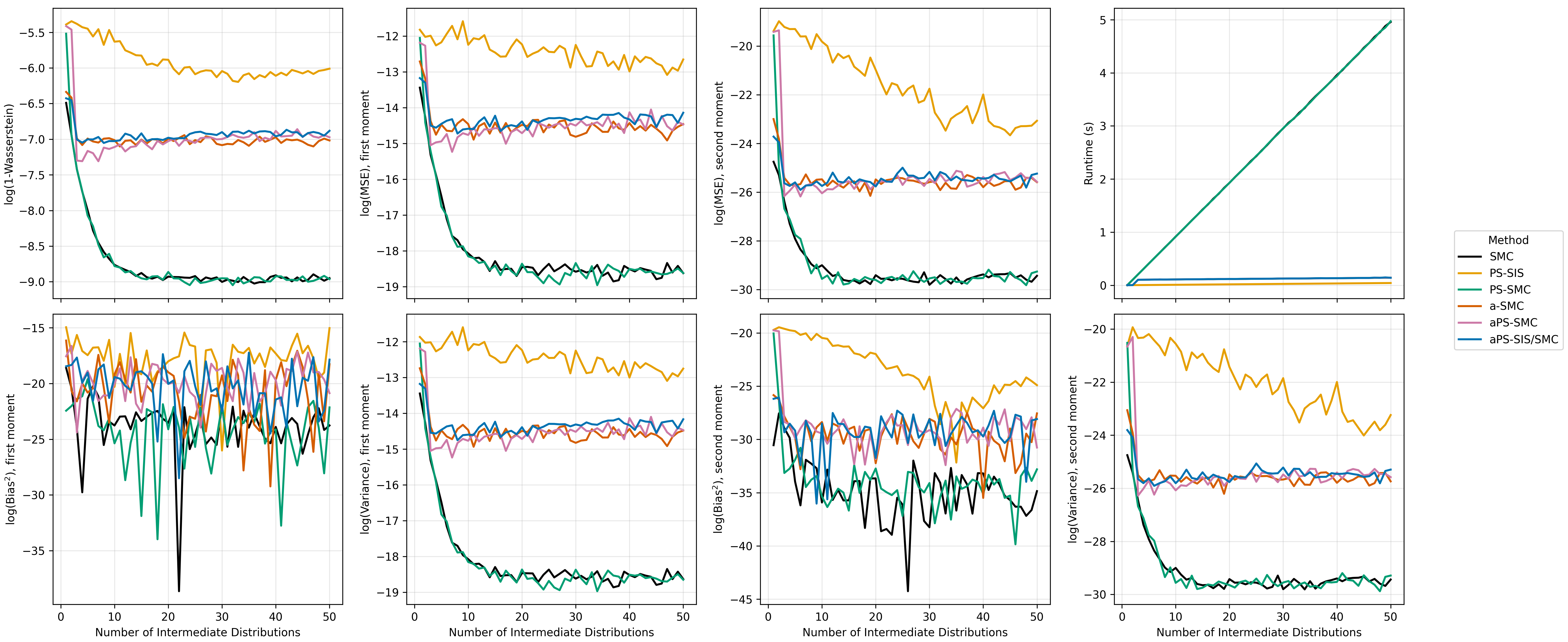}
    \caption{Detailed performance summaries for the variance-contraction
    Gaussian SMC example \(N(0,1)\rightarrow N(0,10^{-5})\). The first row
    shows log Wasserstein distance, first-moment log MSE, second-moment log
    MSE, and runtime. The second row shows first-moment log squared bias,
    first-moment log variance, second-moment log squared bias, and
    second-moment log variance.}
    \label{fig:gaussian_smc_results_appendix_fat_to_thin}
\end{figure}

\begin{figure}
    \centering
    \includegraphics[width=\textwidth]{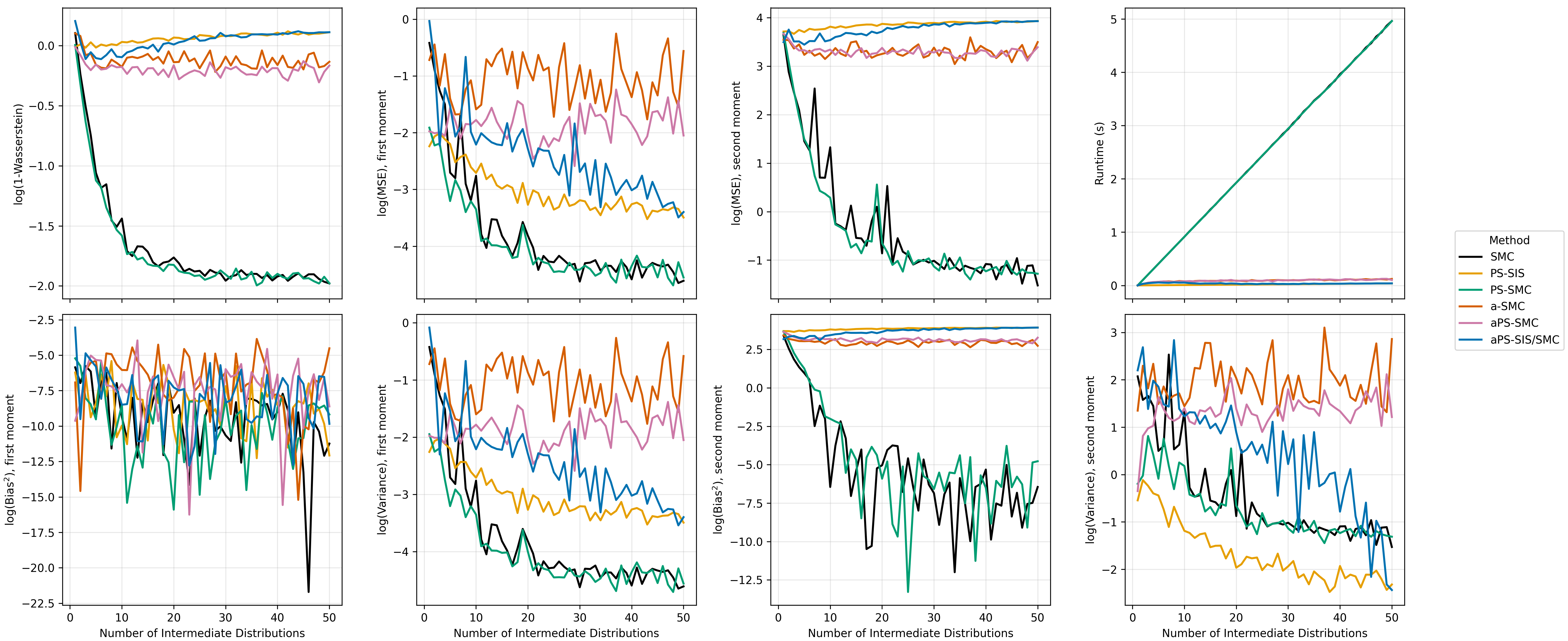}
    \caption{Detailed performance summaries for the variance-expansion
    Gaussian SMC example \(N(0,1)\rightarrow N(0,10)\). The first row shows
    log Wasserstein distance, first-moment log MSE, second-moment log MSE, and
    runtime. The second row shows first-moment log squared bias,
    first-moment log variance, second-moment log squared bias, and
    second-moment log variance.}
    \label{fig:gaussian_smc_results_appendix_thin_to_fat}
\end{figure}

Figures~\ref{fig:gaussian_smc_results_appendix_mean_shift}--\ref{fig:gaussian_smc_results_appendix_thin_to_fat}
give the full decomposition of the moment-estimation errors for the Gaussian
SMC experiments in Section~\ref{sec:ps-smc}. In addition to the Wasserstein
distance, moment MSEs, and runtimes shown in the main text, the figures report
the squared bias and Monte Carlo variance components for both the first and
second moments.

\subsection{Additional g-and-k experiment results}
\label{app:gandk_additional_results}

Tables~\ref{tab:abc_smc_metrics_gandk_means} and
\ref{tab:abc_smc_metrics_gandk_stds} report the repeated-run estimation metrics
for the posterior means and posterior standard deviations in the g-and-k
experiment.

\begin{table}
\caption{Repeated-run ABC-SMC estimation metrics for the \emph{posterior means} of the g-and-k model parameters \(A\), \(B\), \(g\), and \(k\). Bold indicates the smallest value among the non-standard methods.}
\label{tab:abc_smc_metrics_gandk_means}
\begin{tabular*}{\tblwidth}{@{\extracolsep{\fill}}llccc@{}}
\toprule
Parameter & ABC-SMC Method & Mean Squared Error & Variance & Bias\(^2\)  \\ 
\midrule
\multirow{8}{*}{\(A\)} 
  & SMC                    & \(2.95\times10^{-7}\)   & \(2.95\times10^{-7}\)   & \(6.82\times10^{-11}\) \\
  & a-SMC                  & \(1.96\times10^{-4}\)   & \(1.96\times10^{-4}\)   & \(4.93\times10^{-7}\)  \\
  & aPS-SMC                & \(1.56\times10^{-4}\)   & \(1.56\times10^{-4}\)   & \(\mathbf{1.50\times10^{-7}}\) \\
  & aPS-SIS/SMC            & \(9.47\times10^{-5}\)   & \(9.28\times10^{-5}\)   & \(1.88\times10^{-6}\)  \\
  & T-SMC                  & \(\mathbf{3.86\times10^{-5}}\) & \(\mathbf{3.82\times10^{-5}}\) & \(4.12\times10^{-7}\)  \\
  & Jitter, \(c=1\)        & \(1.39\times10^{-2}\)   & \(1.37\times10^{-2}\)   & \(1.83\times10^{-4}\)  \\
  & Jitter, \(c=0.1\)      & \(6.18\times10^{-5}\)   & \(6.14\times10^{-5}\)   & \(4.13\times10^{-7}\)  \\
  & Jitter, \(c=0.01\)     & \(1.24\times10^{-4}\)   & \(1.22\times10^{-4}\)   & \(1.50\times10^{-6}\)  \\
\midrule
\multirow{8}{*}{\(B\)} 
  & SMC                    & \(2.12\times10^{-5}\)   & \(2.02\times10^{-5}\)   & \(9.10\times10^{-7}\)  \\
  & a-SMC                  & \(2.61\times10^{-2}\)   & \(1.82\times10^{-2}\)   & \(7.91\times10^{-3}\)  \\
  & aPS-SMC                & \(1.57\times10^{-2}\)   & \(9.60\times10^{-3}\)   & \(6.07\times10^{-3}\)  \\
  & aPS-SIS/SMC            & \(7.23\times10^{-3}\)   & \(4.59\times10^{-3}\)   & \(\mathbf{2.64\times10^{-3}}\) \\
  & T-SMC                  & \(\mathbf{5.50\times10^{-3}}\) & \(2.85\times10^{-3}\)   & \(2.65\times10^{-3}\)  \\
  & Jitter, \(c=1\)        & \(1.01\times10^{-1}\)   & \(2.86\times10^{-2}\)   & \(7.20\times10^{-2}\)  \\
  & Jitter, \(c=0.1\)      & \(5.61\times10^{-3}\)   & \(\mathbf{2.53\times10^{-3}}\) & \(3.08\times10^{-3}\)  \\
  & Jitter, \(c=0.01\)     & \(1.34\times10^{-2}\)   & \(9.24\times10^{-3}\)   & \(4.17\times10^{-3}\)  \\
\midrule
\multirow{8}{*}{\(g\)} 
  & SMC                    & \(3.45\times10^{-4}\)   & \(3.44\times10^{-4}\)   & \(7.71\times10^{-7}\)  \\
  & a-SMC                  & \(2.19\times10^{-2}\)   & \(2.19\times10^{-2}\)   & \(\mathbf{3.46\times10^{-6}}\) \\
  & aPS-SMC                & \(1.88\times10^{-2}\)   & \(1.86\times10^{-2}\)   & \(1.36\times10^{-4}\)  \\
  & aPS-SIS/SMC            & \(1.00\times10^{-2}\)   & \(1.00\times10^{-2}\)   & \(1.88\times10^{-5}\)  \\
  & T-SMC                  & \(1.84\times10^{-2}\)   & \(1.84\times10^{-2}\)   & \(4.00\times10^{-6}\)  \\
  & Jitter, \(c=1\)        & \(3.26\times10^{-1}\)   & \(3.26\times10^{-1}\)   & \(4.66\times10^{-4}\)  \\
  & Jitter, \(c=0.1\)      & \(\mathbf{7.94\times10^{-3}}\) & \(\mathbf{7.85\times10^{-3}}\) & \(9.23\times10^{-5}\)  \\
  & Jitter, \(c=0.01\)     & \(8.20\times10^{-3}\)   & \(8.20\times10^{-3}\)   & \(7.85\times10^{-6}\)  \\
\midrule
\multirow{8}{*}{\(k\)} 
  & SMC                    & \(1.48\times10^{-3}\)   & \(1.42\times10^{-3}\)   & \(5.92\times10^{-5}\)  \\
  & a-SMC                  & \(3.09\times10^{-1}\)   & \(1.89\times10^{-1}\)   & \(1.20\times10^{-1}\)  \\
  & aPS-SMC                & \(3.05\times10^{-1}\)   & \(2.01\times10^{-1}\)   & \(1.04\times10^{-1}\)  \\
  & aPS-SIS/SMC            & \(1.63\times10^{-1}\)   & \(1.05\times10^{-1}\)   & \(5.76\times10^{-2}\)  \\
  & T-SMC                  & \(2.65\times10^{-1}\)   & \(1.37\times10^{-1}\)   & \(1.28\times10^{-1}\)  \\
  & Jitter, \(c=1\)        & \(5.82\times10^{0}\)    & \(2.88\times10^{0}\)    & \(2.95\times10^{0}\)   \\
  & Jitter, \(c=0.1\)      & \(\mathbf{1.40\times10^{-1}}\) & \(\mathbf{9.28\times10^{-2}}\) & \(\mathbf{4.70\times10^{-2}}\) \\
  & Jitter, \(c=0.01\)     & \(1.99\times10^{-1}\)   & \(1.35\times10^{-1}\)   & \(6.51\times10^{-2}\)  \\
\bottomrule
\end{tabular*}
\end{table}

\begin{table}
\caption{Repeated-run ABC-SMC estimation metrics for the \emph{posterior standard deviations} of the g-and-k model parameters \(A\), \(B\), \(g\), and \(k\). Bold indicates the smallest value among the non-standard methods.}
\label{tab:abc_smc_metrics_gandk_stds}
\begin{tabular*}{\tblwidth}{@{\extracolsep{\fill}}llccc@{}}
\toprule
Parameter & ABC-SMC Method & Mean Squared Error & Variance & Bias\(^2\)  \\ 
\midrule
\multirow{8}{*}{\(A\)} 
  & SMC                    & \(1.00\times10^{-7}\)    & \(1.00\times10^{-7}\)    & \(2.58\times10^{-10}\) \\
  & a-SMC                  & \(1.08\times10^{-5}\)    & \(9.45\times10^{-6}\)    & \(1.37\times10^{-6}\)  \\
  & aPS-SMC                & \(1.80\times10^{-5}\)    & \(1.80\times10^{-5}\)    & \(8.59\times10^{-8}\)  \\
  & aPS-SIS/SMC            & \(8.35\times10^{-6}\)    & \(8.22\times10^{-6}\)    & \(1.27\times10^{-7}\)  \\
  & T-SMC                  & \(8.82\times10^{-6}\)    & \(8.50\times10^{-6}\)    & \(3.18\times10^{-7}\)  \\
  & Jitter, \(c=1\)        & \(3.53\times10^{-2}\)    & \(2.49\times10^{-2}\)    & \(1.04\times10^{-2}\)  \\
  & Jitter, \(c=0.1\)      & \(1.44\times10^{-4}\)    & \(1.31\times10^{-4}\)    & \(1.38\times10^{-5}\)  \\
  & Jitter, \(c=0.01\)     & \(\mathbf{6.14\times10^{-6}}\) & \(\mathbf{6.10\times10^{-6}}\) & \(\mathbf{3.85\times10^{-8}}\) \\
\midrule
\multirow{8}{*}{\(B\)} 
  & SMC                    & \(4.86\times10^{-6}\)    & \(4.73\times10^{-6}\)    & \(1.31\times10^{-7}\)  \\
  & a-SMC                  & \(5.55\times10^{-4}\)    & \(5.37\times10^{-4}\)    & \(1.82\times10^{-5}\)  \\
  & aPS-SMC                & \(5.83\times10^{-4}\)    & \(5.82\times10^{-4}\)    & \(5.10\times10^{-7}\)  \\
  & aPS-SIS/SMC            & \(\mathbf{3.83\times10^{-4}}\) & \(3.81\times10^{-4}\)    & \(1.83\times10^{-6}\)  \\
  & T-SMC                  & \(4.27\times10^{-4}\)    & \(4.10\times10^{-4}\)    & \(1.67\times10^{-5}\)  \\
  & Jitter, \(c=1\)        & \(1.34\times10^{-2}\)    & \(6.27\times10^{-3}\)    & \(7.14\times10^{-3}\)  \\
  & Jitter, \(c=0.1\)      & \(4.56\times10^{-4}\)    & \(\mathbf{3.45\times10^{-4}}\) & \(1.10\times10^{-4}\)  \\
  & Jitter, \(c=0.01\)     & \(4.68\times10^{-4}\)    & \(4.68\times10^{-4}\)    & \(\mathbf{1.50\times10^{-7}}\) \\
\midrule
\multirow{8}{*}{\(g\)} 
  & SMC                    & \(3.51\times10^{-4}\)    & \(2.92\times10^{-4}\)    & \(5.94\times10^{-5}\)  \\
  & a-SMC                  & \(7.65\times10^{-3}\)    & \(2.48\times10^{-3}\)    & \(5.17\times10^{-3}\)  \\
  & aPS-SMC                & \(5.44\times10^{-3}\)    & \(2.13\times10^{-3}\)    & \(3.32\times10^{-3}\)  \\
  & aPS-SIS/SMC            & \(4.59\times10^{-3}\)    & \(1.43\times10^{-3}\)    & \(3.17\times10^{-3}\)  \\
  & T-SMC                  & \(\mathbf{3.79\times10^{-3}}\) & \(1.69\times10^{-3}\)    & \(2.11\times10^{-3}\)  \\
  & Jitter, \(c=1\)        & \(9.42\times10^{-1}\)    & \(6.71\times10^{-1}\)    & \(2.71\times10^{-1}\)  \\
  & Jitter, \(c=0.1\)      & \(3.90\times10^{-3}\)    & \(2.31\times10^{-3}\)    & \(\mathbf{1.60\times10^{-3}}\) \\
  & Jitter, \(c=0.01\)     & \(5.03\times10^{-3}\)    & \(\mathbf{1.03\times10^{-3}}\) & \(3.99\times10^{-3}\)  \\
\midrule
\multirow{8}{*}{\(k\)} 
  & SMC                    & \(3.12\times10^{-4}\)    & \(3.11\times10^{-4}\)    & \(1.18\times10^{-6}\)  \\
  & a-SMC                  & \(2.63\times10^{-2}\)    & \(1.88\times10^{-2}\)    & \(7.57\times10^{-3}\)  \\
  & aPS-SMC                & \(1.75\times10^{-2}\)    & \(1.36\times10^{-2}\)    & \(3.93\times10^{-3}\)  \\
  & aPS-SIS/SMC            & \(1.64\times10^{-2}\)    & \(1.52\times10^{-2}\)    & \(1.27\times10^{-3}\)  \\
  & T-SMC                  & \(1.85\times10^{-2}\)    & \(1.85\times10^{-2}\)    & \(\mathbf{4.60\times10^{-5}}\) \\
  & Jitter, \(c=1\)        & \(9.06\times10^{-1}\)    & \(6.10\times10^{-1}\)    & \(2.96\times10^{-1}\)  \\
  & Jitter, \(c=0.1\)      & \(\mathbf{8.59\times10^{-3}}\) & \(\mathbf{8.24\times10^{-3}}\) & \(3.45\times10^{-4}\)  \\
  & Jitter, \(c=0.01\)     & \(1.57\times10^{-2}\)    & \(1.17\times10^{-2}\)    & \(3.97\times10^{-3}\)  \\
\bottomrule
\end{tabular*}
\end{table}

\subsection{Additional Lotka--Volterra experiment results}
\label{app:lv_additional_results}

Tables~\ref{tab:abc_smc_metrics_lv_means_app} and
\ref{tab:abc_smc_metrics_lv_sds_app} report the repeated-run estimation
metrics for the posterior means and posterior standard deviations in the
Lotka--Volterra experiment.

\begin{table}
\caption{Repeated-run ABC-SMC estimation metrics for the \emph{posterior means} of the Lotka--Volterra parameters \(k_1\), \(k_2\), and \(k_3\). Bold indicates the smallest value among the non-standard methods.}
\label{tab:abc_smc_metrics_lv_means_app}
\begin{tabular*}{\tblwidth}{@{\extracolsep{\fill}}llccc@{}}
\toprule
Parameter & ABC-SMC Method & Mean Squared Error & Variance & Bias\(^2\) \\
\midrule
\multirow{8}{*}{\(k_1\)}
  & SMC                    & \(2.63\times10^{-3}\) & \(2.31\times10^{-3}\) & \(3.18\times10^{-4}\) \\
  & a-SMC                  & \(2.83\times10^{-1}\) & \(2.73\times10^{-1}\) & \(9.34\times10^{-3}\) \\
  & aPS-SMC                & \(2.98\times10^{-1}\) & \(2.89\times10^{-1}\) & \(8.65\times10^{-3}\) \\
  & aPS-SIS/SMC            & \(\mathbf{1.80\times10^{-1}}\) & \(\mathbf{1.76\times10^{-1}}\) & \(4.03\times10^{-3}\) \\
  & T-SMC                  & \(2.06\times10^{-1}\) & \(2.04\times10^{-1}\) & \(\mathbf{1.68\times10^{-3}}\) \\
  & Jitter, \(c=1\)        & \(1.31\times10^{1}\)  & \(1.38\times10^{0}\)  & \(1.17\times10^{1}\) \\
  & Jitter, \(c=0.1\)      & \(1.18\times10^{1}\)  & \(1.86\times10^{0}\)  & \(9.93\times10^{0}\) \\
  & Jitter, \(c=0.01\)     & \(1.01\times10^{1}\)  & \(3.48\times10^{0}\)  & \(6.57\times10^{0}\) \\
\midrule
\multirow{8}{*}{\(k_2\)}
  & SMC                    & \(1.02\times10^{-7}\) & \(8.61\times10^{-8}\) & \(1.63\times10^{-8}\) \\
  & a-SMC                  & \(2.15\times10^{-5}\) & \(1.99\times10^{-5}\) & \(1.57\times10^{-6}\) \\
  & aPS-SMC                & \(9.69\times10^{-6}\) & \(8.48\times10^{-6}\) & \(1.21\times10^{-6}\) \\
  & aPS-SIS/SMC            & \(1.54\times10^{-5}\) & \(1.44\times10^{-5}\) & \(9.33\times10^{-7}\) \\
  & T-SMC                  & \(\mathbf{2.13\times10^{-6}}\) & \(\mathbf{1.83\times10^{-6}}\) & \(\mathbf{2.93\times10^{-7}}\) \\
  & Jitter, \(c=1\)        & \(2.59\times10^{0}\)  & \(2.01\times10^{0}\)  & \(5.79\times10^{-1}\) \\
  & Jitter, \(c=0.1\)      & \(4.83\times10^{-4}\) & \(4.82\times10^{-4}\) & \(9.91\times10^{-7}\) \\
  & Jitter, \(c=0.01\)     & \(3.48\times10^{-6}\) & \(2.00\times10^{-6}\) & \(1.48\times10^{-6}\) \\
\midrule
\multirow{8}{*}{\(k_3\)}
  & SMC                    & \(2.80\times10^{-3}\) & \(2.74\times10^{-3}\) & \(6.35\times10^{-5}\) \\
  & a-SMC                  & \(3.64\times10^{-1}\) & \(3.34\times10^{-1}\) & \(2.95\times10^{-2}\) \\
  & aPS-SMC                & \(1.89\times10^{-1}\) & \(1.57\times10^{-1}\) & \(3.23\times10^{-2}\) \\
  & aPS-SIS/SMC            & \(3.16\times10^{-1}\) & \(2.89\times10^{-1}\) & \(2.71\times10^{-2}\) \\
  & T-SMC                  & \(\mathbf{9.88\times10^{-2}}\) & \(\mathbf{8.74\times10^{-2}}\) & \(\mathbf{1.13\times10^{-2}}\) \\
  & Jitter, \(c=1\)        & \(1.23\times10^{1}\)  & \(1.47\times10^{0}\)  & \(1.09\times10^{1}\) \\
  & Jitter, \(c=0.1\)      & \(1.16\times10^{1}\)  & \(2.02\times10^{0}\)  & \(9.57\times10^{0}\) \\
  & Jitter, \(c=0.01\)     & \(8.64\times10^{0}\)  & \(2.78\times10^{0}\)  & \(5.86\times10^{0}\) \\
\bottomrule
\end{tabular*}
\end{table}

\begin{table}
\caption{Repeated-run ABC-SMC estimation metrics for the \emph{posterior standard deviations} of the Lotka--Volterra parameters \(k_1\), \(k_2\), and \(k_3\). Bold indicates the smallest value among the non-standard methods.}
\label{tab:abc_smc_metrics_lv_sds_app}
\begin{tabular*}{\tblwidth}{@{\extracolsep{\fill}}llccc@{}}
\toprule
Parameter & ABC-SMC Method & Mean Squared Error & Variance & Bias\(^2\) \\
\midrule
\multirow{8}{*}{\(k_1\)}
  & SMC                    & \(7.37\times10^{-3}\) & \(7.07\times10^{-3}\) & \(2.95\times10^{-4}\) \\
  & a-SMC                  & \(1.44\times10^{-2}\) & \(1.19\times10^{-2}\) & \(2.52\times10^{-3}\) \\
  & aPS-SMC                & \(3.98\times10^{-2}\) & \(3.93\times10^{-2}\) & \(5.03\times10^{-4}\) \\
  & aPS-SIS/SMC            & \(2.57\times10^{-2}\) & \(2.55\times10^{-2}\) & \(\mathbf{2.61\times10^{-4}}\) \\
  & T-SMC                  & \(\mathbf{4.07\times10^{-3}}\) & \(\mathbf{1.72\times10^{-3}}\) & \(2.35\times10^{-3}\) \\
  & Jitter, \(c=1\)        & \(1.05\times10^{0}\)  & \(3.99\times10^{-1}\) & \(6.47\times10^{-1}\) \\
  & Jitter, \(c=0.1\)      & \(2.00\times10^{-1}\) & \(1.18\times10^{-1}\) & \(8.17\times10^{-2}\) \\
  & Jitter, \(c=0.01\)     & \(1.86\times10^{-1}\) & \(1.27\times10^{-1}\) & \(5.89\times10^{-2}\) \\
\midrule
\multirow{8}{*}{\(k_2\)}
  & SMC                    & \(2.73\times10^{-7}\) & \(2.64\times10^{-7}\) & \(8.95\times10^{-9}\) \\
  & a-SMC                  & \(1.25\times10^{-6}\) & \(1.23\times10^{-6}\) & \(1.55\times10^{-8}\) \\
  & aPS-SMC                & \(8.33\times10^{-7}\) & \(8.33\times10^{-7}\) & \(\mathbf{1.22\times10^{-11}}\) \\
  & aPS-SIS/SMC            & \(7.05\times10^{-7}\) & \(7.04\times10^{-7}\) & \(6.59\times10^{-10}\) \\
  & T-SMC                  & \(\mathbf{6.93\times10^{-8}}\) & \(\mathbf{4.84\times10^{-8}}\) & \(2.10\times10^{-8}\) \\
  & Jitter, \(c=1\)        & \(6.48\times10^{-1}\) & \(5.20\times10^{-1}\) & \(1.28\times10^{-1}\) \\
  & Jitter, \(c=0.1\)      & \(9.68\times10^{-7}\) & \(9.67\times10^{-7}\) & \(7.63\times10^{-10}\) \\
  & Jitter, \(c=0.01\)     & \(8.47\times10^{-7}\) & \(8.35\times10^{-7}\) & \(1.19\times10^{-8}\) \\
\midrule
\multirow{8}{*}{\(k_3\)}
  & SMC                    & \(8.33\times10^{-3}\) & \(8.05\times10^{-3}\) & \(2.77\times10^{-4}\) \\
  & a-SMC                  & \(2.77\times10^{-2}\) & \(2.75\times10^{-2}\) & \(2.40\times10^{-4}\) \\
  & aPS-SMC                & \(2.13\times10^{-2}\) & \(2.13\times10^{-2}\) & \(\mathbf{6.98\times10^{-6}}\) \\
  & aPS-SIS/SMC            & \(1.84\times10^{-2}\) & \(1.83\times10^{-2}\) & \(1.01\times10^{-4}\) \\
  & T-SMC                  & \(\mathbf{1.46\times10^{-3}}\) & \(\mathbf{9.87\times10^{-4}}\) & \(4.75\times10^{-4}\) \\
  & Jitter, \(c=1\)        & \(1.13\times10^{0}\)  & \(4.12\times10^{-1}\) & \(7.17\times10^{-1}\) \\
  & Jitter, \(c=0.1\)      & \(1.81\times10^{-1}\) & \(9.39\times10^{-2}\) & \(8.75\times10^{-2}\) \\
  & Jitter, \(c=0.01\)     & \(1.91\times10^{-1}\) & \(1.18\times10^{-1}\) & \(7.25\times10^{-2}\) \\
\bottomrule
\end{tabular*}
\end{table}

\subsection{Additional Ricker experiment results}
\label{app:ricker_additional_results}

Tables~\ref{tab:abc_smc_metrics_ricker_means_app} and
\ref{tab:abc_smc_metrics_ricker_stds_app} report the repeated-run estimation
metrics for the posterior means and posterior standard deviations in the Ricker
experiment.

\begin{table}
\caption{Repeated-run ABC-SMC estimation metrics for the \emph{posterior means} of the Ricker model parameters \(\log R\), \(\log \sigma\), and \(\log \phi\). Bold indicates the smallest value among the non-standard methods.}
\label{tab:abc_smc_metrics_ricker_means_app}
\begin{tabular*}{\tblwidth}{@{\extracolsep{\fill}}llccc@{}}
\toprule
Parameter & ABC-SMC Method & Mean Squared Error & Variance & Bias\(^2\) \\
\midrule
\multirow{8}{*}{\(\log R\)} 
  & SMC               & \(1.34\times10^{-2}\) & \(1.22\times10^{-2}\) & \(1.23\times10^{-3}\) \\
  & a-SMC             & \(3.00\times10^{-2}\) & \(2.86\times10^{-2}\) & \(1.46\times10^{-3}\) \\
  & aPS-SMC           & \(2.27\times10^{-2}\) & \(7.31\times10^{-3}\) & \(1.54\times10^{-2}\) \\
  & aPS-SIS/SMC       & \(2.22\times10^{-2}\) & \(1.95\times10^{-2}\) & \(2.73\times10^{-3}\) \\
  & T-SMC             & \(2.15\times10^{-2}\) & \(7.94\times10^{-3}\) & \(1.36\times10^{-2}\) \\
  & Jitter, \(c=1\)   & \(3.74\times10^{-2}\) & \(3.63\times10^{-2}\) & \(\mathbf{1.11\times10^{-3}}\) \\
  & Jitter, \(c=0.1\) & \(\mathbf{7.02\times10^{-3}}\) & \(\mathbf{1.24\times10^{-3}}\) & \(5.78\times10^{-3}\) \\
  & Jitter, \(c=0.01\)& \(1.40\times10^{-2}\) & \(1.90\times10^{-3}\) & \(1.21\times10^{-2}\) \\
\midrule
\multirow{8}{*}{\(\log \sigma\)} 
  & SMC               & \(1.26\times10^{-2}\) & \(1.17\times10^{-2}\) & \(8.96\times10^{-4}\) \\
  & a-SMC             & \(2.58\times10^{-2}\) & \(2.56\times10^{-2}\) & \(1.82\times10^{-4}\) \\
  & aPS-SMC           & \(2.37\times10^{-2}\) & \(2.36\times10^{-2}\) & \(5.64\times10^{-5}\) \\
  & aPS-SIS/SMC       & \(1.68\times10^{-2}\) & \(1.67\times10^{-2}\) & \(8.25\times10^{-5}\) \\
  & T-SMC             & \(1.68\times10^{-2}\) & \(1.68\times10^{-2}\) & \(2.11\times10^{-5}\) \\
  & Jitter, \(c=1\)   & \(\mathbf{3.88\times10^{-3}}\) & \(\mathbf{3.84\times10^{-3}}\) & \(4.20\times10^{-5}\) \\
  & Jitter, \(c=0.1\) & \(1.06\times10^{-2}\) & \(9.14\times10^{-3}\) & \(1.45\times10^{-3}\) \\
  & Jitter, \(c=0.01\)& \(1.25\times10^{-2}\) & \(1.25\times10^{-2}\) & \(\mathbf{6.00\times10^{-6}}\) \\
\midrule
\multirow{8}{*}{\(\log \phi\)} 
  & SMC               & \(2.92\times10^{-3}\) & \(2.56\times10^{-3}\) & \(3.60\times10^{-4}\) \\
  & a-SMC             & \(8.52\times10^{-3}\) & \(7.88\times10^{-3}\) & \(6.45\times10^{-4}\) \\
  & aPS-SMC           & \(1.85\times10^{-2}\) & \(6.66\times10^{-3}\) & \(1.18\times10^{-2}\) \\
  & aPS-SIS/SMC       & \(8.00\times10^{-3}\) & \(6.87\times10^{-3}\) & \(1.14\times10^{-3}\) \\
  & T-SMC             & \(5.46\times10^{-3}\) & \(5.13\times10^{-3}\) & \(\mathbf{3.26\times10^{-4}}\) \\
  & Jitter, \(c=1\)   & \(5.04\times10^{-2}\) & \(2.43\times10^{-2}\) & \(2.61\times10^{-2}\) \\
  & Jitter, \(c=0.1\) & \(\mathbf{3.66\times10^{-3}}\) & \(\mathbf{1.47\times10^{-3}}\) & \(2.19\times10^{-3}\) \\
  & Jitter, \(c=0.01\)& \(4.74\times10^{-3}\) & \(1.91\times10^{-3}\) & \(2.83\times10^{-3}\) \\
\bottomrule
\end{tabular*}
\end{table}

\begin{table}
\caption{Repeated-run ABC-SMC estimation metrics for the \emph{posterior standard deviations} of the Ricker model parameters \(\log R\), \(\log \sigma\), and \(\log \phi\). Bold indicates the smallest value among the non-standard methods.}
\label{tab:abc_smc_metrics_ricker_stds_app}
\begin{tabular*}{\tblwidth}{@{\extracolsep{\fill}}llccc@{}}
\toprule
Parameter & ABC-SMC Method & Mean Squared Error & Variance & Bias\(^2\) \\
\midrule
\multirow{8}{*}{\(\log R\)} 
  & SMC               & \(4.70\times10^{-3}\) & \(1.91\times10^{-3}\) & \(2.79\times10^{-3}\) \\
  & a-SMC             & \(4.38\times10^{-3}\) & \(4.38\times10^{-3}\) & \(\mathbf{3.01\times10^{-8}}\) \\
  & aPS-SMC           & \(2.63\times10^{-3}\) & \(2.03\times10^{-3}\) & \(5.95\times10^{-4}\) \\
  & aPS-SIS/SMC       & \(5.87\times10^{-3}\) & \(3.77\times10^{-3}\) & \(2.10\times10^{-3}\) \\
  & T-SMC             & \(2.34\times10^{-3}\) & \(2.24\times10^{-3}\) & \(1.02\times10^{-4}\) \\
  & Jitter, \(c=1\)   & \(1.87\times10^{-1}\) & \(3.19\times10^{-2}\) & \(1.55\times10^{-1}\) \\
  & Jitter, \(c=0.1\) & \(\mathbf{1.04\times10^{-3}}\) & \(\mathbf{4.49\times10^{-4}}\) & \(5.92\times10^{-4}\) \\
  & Jitter, \(c=0.01\)& \(1.95\times10^{-3}\) & \(5.21\times10^{-4}\) & \(1.43\times10^{-3}\) \\
\midrule
\multirow{8}{*}{\(\log \sigma\)} 
  & SMC               & \(4.74\times10^{-3}\) & \(2.69\times10^{-3}\) & \(2.05\times10^{-3}\) \\
  & a-SMC             & \(1.64\times10^{-2}\) & \(4.21\times10^{-3}\) & \(1.22\times10^{-2}\) \\
  & aPS-SMC           & \(1.19\times10^{-2}\) & \(4.11\times10^{-3}\) & \(7.77\times10^{-3}\) \\
  & aPS-SIS/SMC       & \(7.46\times10^{-3}\) & \(2.77\times10^{-3}\) & \(4.69\times10^{-3}\) \\
  & T-SMC             & \(9.26\times10^{-3}\) & \(4.25\times10^{-3}\) & \(5.01\times10^{-3}\) \\
  & Jitter, \(c=1\)   & \(\mathbf{1.88\times10^{-3}}\) & \(\mathbf{9.00\times10^{-4}}\) & \(\mathbf{9.83\times10^{-4}}\) \\
  & Jitter, \(c=0.1\) & \(4.41\times10^{-3}\) & \(1.53\times10^{-3}\) & \(2.88\times10^{-3}\) \\
  & Jitter, \(c=0.01\)& \(7.96\times10^{-3}\) & \(2.85\times10^{-3}\) & \(5.11\times10^{-3}\) \\
\midrule
\multirow{8}{*}{\(\log \phi\)} 
  & SMC               & \(6.54\times10^{-4}\) & \(5.91\times10^{-4}\) & \(6.28\times10^{-5}\) \\
  & a-SMC             & \(2.40\times10^{-3}\) & \(1.55\times10^{-3}\) & \(8.48\times10^{-4}\) \\
  & aPS-SMC           & \(1.74\times10^{-3}\) & \(1.39\times10^{-3}\) & \(3.49\times10^{-4}\) \\
  & aPS-SIS/SMC       & \(1.56\times10^{-3}\) & \(1.54\times10^{-3}\) & \(\mathbf{2.71\times10^{-5}}\) \\
  & T-SMC             & \(1.41\times10^{-3}\) & \(1.10\times10^{-3}\) & \(3.14\times10^{-4}\) \\
  & Jitter, \(c=1\)   & \(1.42\times10^{-1}\) & \(3.54\times10^{-2}\) & \(1.06\times10^{-1}\) \\
  & Jitter, \(c=0.1\) & \(\mathbf{5.71\times10^{-4}}\) & \(\mathbf{4.62\times10^{-4}}\) & \(1.09\times10^{-4}\) \\
  & Jitter, \(c=0.01\)& \(1.09\times10^{-3}\) & \(5.77\times10^{-4}\) & \(5.08\times10^{-4}\) \\
\bottomrule
\end{tabular*}
\end{table}

\section{Tempering schedule for Gaussian examples}
\label{app:gaussian_tempering_schedule}

This appendix gives the derivation of the closed-form tempering schedule used
for the one-dimensional Gaussian examples in Section~\ref{sec:ps-smc}. Consider the tempered path
\[
\pi_{\lambda}(\theta)
\propto
\pi_0(\theta)^{1-\lambda}
\pi_T(\theta)^{\lambda},
\qquad \lambda\in[0,1].
\]
\citet{chopin2024tempering} use an approach inspired by mirror descent to choose the tempering schedule. Whilst the argument in that paper is based on a high-dimensional setting, we still find the resultant sequences of targets of use in our one-dimensional examples. In particular, following equation (16) of \citet{chopin2024tempering}, we replace the discrete sequence by a smooth schedule \(\lambda(t)\), \(t\in[0,1]\), satisfying
\[
\lambda(0)=0,
\qquad
\lambda(1)=1,
\]
and solve an ordinary differential equation of the form
\[
\dot{\lambda}=c I(\lambda)^{-1/2},
\]
where \(I(\lambda)\) is the Fisher-information-type quantity along the tempering path and \(c>0\) is chosen so that \(\lambda(1)=1\).

First consider the variance-change case
\[
\pi_0=N(0,v_0),
\qquad
\pi_T=N(0,v_T).
\]
For a homogeneous Gaussian variance change, the resulting ordinary differential
equation can be written as
\[
\frac{d\lambda}{dt}
=
c(1+a\lambda),
\]
where
\[
a=\frac{v_0}{v_T}-1
\]
and \(c>0\) is chosen so that the boundary condition \(\lambda(1)=1\) is
satisfied. This is the schedule used for the variance-contraction example
\(N(0,1)\rightarrow N(0,10^{-5})\) and the variance-expansion example
\(N(0,1)\rightarrow N(0,10)\).

Solving the ODE gives
\[
\frac{d\lambda}{1+a\lambda}=c\,dt.
\]
Integrating both sides,
\[
\int \frac{1}{1+a\lambda}\,d\lambda
=
\int c\,dt,
\]
and therefore
\[
\frac{1}{a}\log(1+a\lambda)=ct+C.
\]
Using \(\lambda(0)=0\) gives \(C=0\), so
\[
\frac{1}{a}\log(1+a\lambda)=ct.
\]
Hence
\[
1+a\lambda=\exp(act),
\]
and
\[
\lambda(t)=\frac{\exp(act)-1}{a}.
\]
The condition \(\lambda(1)=1\) implies
\[
1=\frac{\exp(ac)-1}{a},
\]
so that
\[
\exp(ac)=1+a.
\]
Thus
\[
ac=\log(1+a),
\]
and substituting this into the solution gives
\[
\lambda(t)
=
\frac{\exp\{t\log(1+a)\}-1}{a}
=
\frac{(1+a)^t-1}{a}.
\]
Since \(1+a=v_0/v_T\), the schedule can also be written as
\[
\lambda(t)
=
\frac{(v_0/v_T)^t-1}{(v_0/v_T)-1}.
\]
For \(T\) intermediate steps, we evaluate this expression at
\(t=j/T\), \(j=0,\ldots,T\), giving the schedule
\[
\alpha_j
=
\lambda(j/T)
=
\frac{(v_0/v_T)^{j/T}-1}{(v_0/v_T)-1},
\qquad
j=0,\ldots,T.
\]
When \(v_0=v_T\), the denominator tends to zero and the limiting schedule is
the linear schedule
\[
\alpha_j=\frac{j}{T}.
\]

For the mean-shift example,
\[
\pi_0=N(0,1),
\qquad
\pi_T=N(10,1),
\]
the proposal and target have the same variance and different means. In this
case, the corresponding quantity \(I(\lambda)\) is constant along the path, so
equation (16) of \citet{chopin2024tempering} gives
\[
\frac{d\lambda}{dt}=c.
\]
Together with \(\lambda(0)=0\) and \(\lambda(1)=1\), this gives \(c=1\) and hence
\[
\lambda(t)=t.
\]
Therefore, for \(T\) intermediate steps, the schedule used in the mean-shift
example is simply the linear schedule
\[
\alpha_j=\frac{j}{T},
\qquad
j=0,\ldots,T.
\]

\clearpage

\bibliographystyle{plainnat}
\bibliography{main}

\end{document}